\documentclass{amsart}

\usepackage{microtype}

\usepackage{amsmath}
\usepackage{amsfonts}
\usepackage{amssymb}
\usepackage{cite}
\usepackage{graphicx}
\usepackage{ifpdf}
\usepackage{hyperref}

\newcommand{\shortcite}[1]{\cite{#1}}

\usepackage{etoolbox}

\usepackage[usenames,dvipsnames,table]{xcolor}
\usepackage[export]{adjustbox}
\usepackage[normalem]{ulem}

\newcommand{\enquote}[1]{``#1''}

\newcommand{\myincludegraphics}[2][]{\includegraphics[clip,#1]{images/#2}}

\newcommand{\reals}{\ensuremath{\mathbb{R}}}

\newcommand{\bv}[1]{\mathbf{#1}}
\newcommand{\vecx}{\mathbf{x}}
\newcommand{\vecy}{\mathbf{y}}

\newcommand{\vecu}{\mathbf{u}}
\newcommand{\vecv}{\mathbf{v}}

\newcommand{\vect}{\mathbf{t}}

\newcommand{\prob}{\operatorname{Pr}}

\newcommand{\rot}{\mathbf{R}}
\newcommand{\meanShape}{{b_0}}
\newcommand{\basisShape}[1]{{b_{#1}}}

\newcommand{\urshape}{\mathcal{U}}
\newcommand{\urmesh}{\mathbf{U}}
\newcommand{\shapeParams}{\vec{\lambda}}
\newcommand{\shapeParam}{\lambda}
\newcommand{\shapeFun}{f}
\newcommand{\sspaceDim}{d}
\newcommand{\numVertices}{n}
\newcommand{\shapeFunArgs}[2]{\shapeFun_{#2}(#1)}
\newcommand{\stdDev}{\sigma}
\newcommand{\stdDevs}{\vec{\sigma}}

\newcommand{\allCorr}{\mathcal{X}}
\newcommand{\corr}{x}

\newcommand{\boundarysym}{b}
\newcommand{\partMesh}{\bv{P}}

\title{Compact Part-Based Shape Spaces for Dense Correspondences}

\author[Burghard et al.]{
  Oliver Burghard, University of Bonn \\
  Alexander Berner, University of Bonn \\
  Michael Wand, Saarland University, MPI Informatik\\ $\quad$
  Niloy Mitra, University College, London \\
  Hans-Peter Seidel, MPI Informatik \\
  Reinhard Klein, University of Bonn }


\begin{document}

\begin{abstract}
  We consider the problem of establishing dense correspondences within a set of related shapes of strongly varying geometry. For such input, traditional shape matching approaches often produce unsatisfactory results.
  We propose an ensemble optimization method that improves given coarse correspondences to obtain dense correspondences. Following ideas from minimum description length approaches, it maximizes the compactness of the induced shape space to obtain high-quality correspondences. We make a number of improvements that are important for computer graphics applications: Our approach handles meshes of general topology and handles partial matching between input of varying topology. To this end we introduce a novel part-based generative statistical shape model. We develop a novel analysis algorithm that learns such models from training shapes of varying topology. We also provide a novel synthesis method that can generate new instances with varying part layouts and subject to generic variational constraints.
  In practical experiments, we obtain a substantial improvement in correspondence quality over state-of-the-art methods. As example application, we demonstrate a system that learns shape families as assemblies of deformable parts and permits real-time editing with continuous and discrete variability.
\end{abstract}

\maketitle

\begin{figure}[h!]
  \centering
  \renewcommand{\tabcolsep}{.07em}  \begin{tabular}{ccc}
    \hspace{-0.01\textwidth}
    \myincludegraphics[width=0.33\textwidth]
    {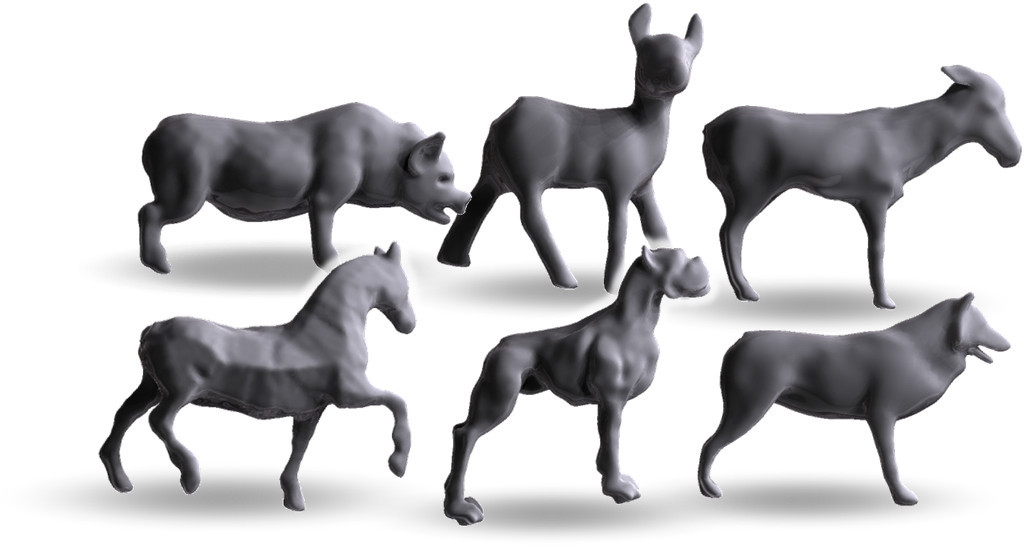} &
    \hspace{-0.01\textwidth}
    \myincludegraphics[width=0.33\textwidth]
    {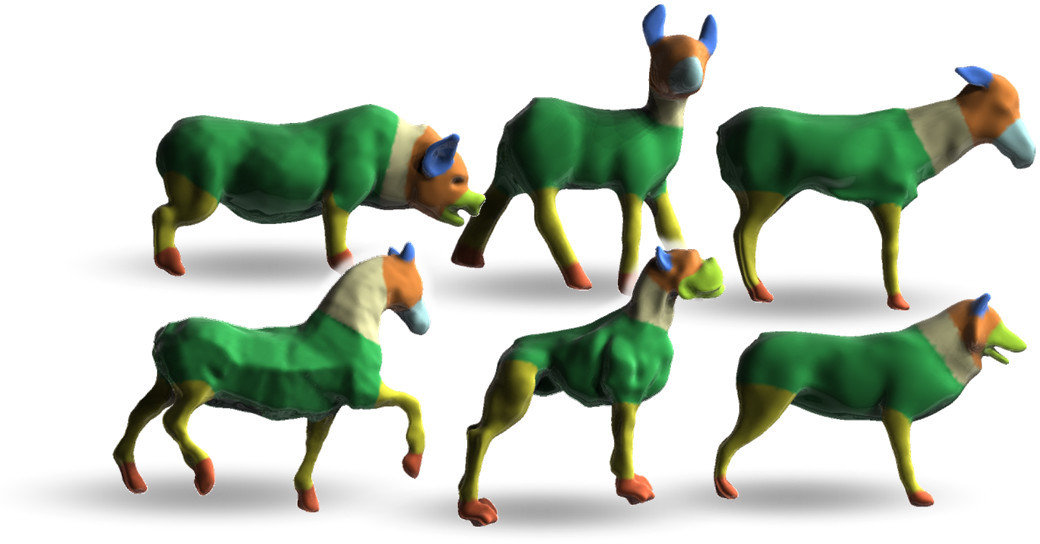} &
    \hspace{-0.01\textwidth}
    \myincludegraphics[width=0.33\textwidth]
    {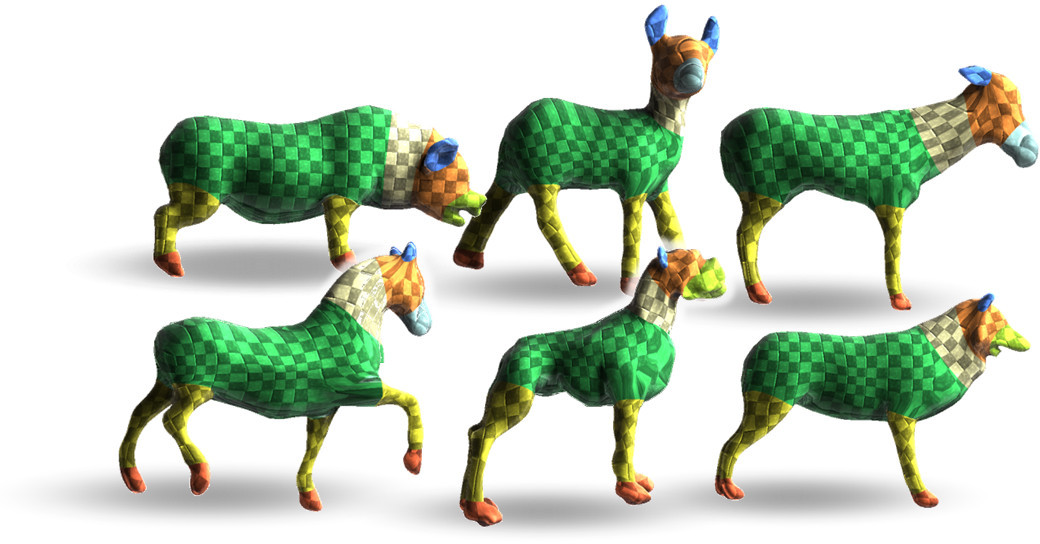}
    \\
    \small (a) shape collection &
    \small (b) coarse segmentation &
    \small (c) dense correspondences
  \end{tabular}
\end{figure}

\begin{figure}
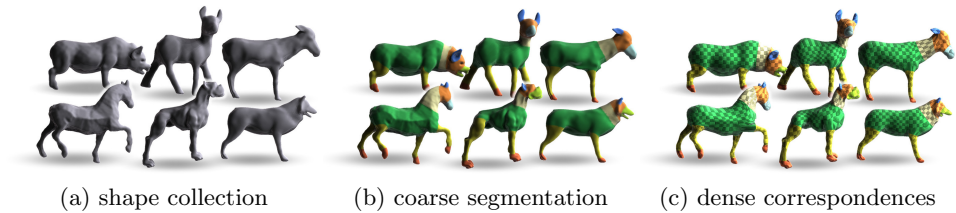

  \centering
  \renewcommand{\tabcolsep}{.07em}
  \begin{tabular}{ccc}
    \hspace{-0.01\textwidth}
    \myincludegraphics[width=0.33\textwidth]
    {TeaserNew1-trim.jpeg} &
    \hspace{-0.01\textwidth}
    \myincludegraphics[width=0.33\textwidth]
    {TeaserNew2-trim.jpeg} &
    \hspace{-0.01\textwidth}
    \myincludegraphics[width=0.33\textwidth]
    {TeaserNew3-trim.jpeg}
    \\
    \small (a) shape collection &
    \small (b) coarse segmentation &
    \small (c) dense correspondences
  \end{tabular}
  \caption[Example of ensemble entropy optimization inputs and outputs]{We compute high-quality dense correspondences between semantically models with strongly varying geometry, such as the four-legged animals in  (a) (6 out of 12 input shapes shown). We require a coarse initialization as input (b) and refine them by maximizing the compactness of a part-wise Gaussian generative model (c).}
  \label{fig:teaser}
\end{figure}

\newcommand{\entropyvideourl}{https://www.youtube.com/watch?v=2m3TbGO9Kls}
\newcommand{\videoWithFootnote}[1]{\href{\entropyvideourl}{#1}\footnote{\url{\entropyvideourl}}}

\section{Introduction}

Computer graphics has reached impressively high standards in representation and rendering of 3D scenes, regularly achieving photo-realism. As a consequence, the problem of creating 3D models of matching quality has become a serious problem, making content creation a major bottleneck in practice.

Data-driven methods are a promising avenue towards addressing this problem. The reuse of existing content, such as models available in large online data-bases, might become a viable option for reducing the content creation costs in the future. In order to be useful as a creative tool, the goal is not to just copy existing models like clip-arts, but to be able to maneuver within the space spanned by the examples and synthesize new shapes of related structure.

An important low-level problem for building such navigatable shape spaces is the correspondence problem: We need to determine which parts of objects are equivalent and which surface points have to be matched, establishing dense correspondences. Most statistical analysis techniques for building parametrized shape spaces require such a dense prior alignment as input \cite{Blanz1999,Allen2003,Hasler2009}. The results crucially depend on the quality of the correspondences: Inaccurate and drifting correspondences yield bad shape spaces. In such spaces, sampling and interpolation yields implausible results (see the accompanying \videoWithFootnote\ for a visualization). In other words, such models fail to \emph{generalize} beyond the input data. The problem can be reduced by using large training sets to learn rather low-dimensional shape spaces. This averages out drift but also reduces the accuracy; only low frequency bands of the geometry are still predicted (see for example the results in \cite{Hasler2009} obtained from almost 2000 shapes).

Hence, good correspondences are a key requirement for building useful and informative shape spaces. The correspondence problem comes in two flavors: \emph{Global} and \emph{local} matching. Local matching requires a rough initialization and refines it but is prone to getting stuck in local optima. Global methods are complementary: They aim at providing the initialization, but are usually unable to compute detailed and dense solutions.
Our paper addresses the local problem: We want to find good dense correspondences given a coarse initialization. Specifically, we address this problem for the case of shapes that have strongly varying geometry, as typically required for learning shape spaces.

Computing correspondences among rather \enquote{similar} shapes is a problem that is by now already quite well understood. Variants of the ICP algorithm handle local alignment \cite{Rusinkiewicz2001} for both rigid and deformable models. Deformable ICP employs differential deformation priors such as elasticity \cite{Haehnel2003}, isometry \cite{Bronstein2006,Ovsjanikov2010}, conformal maps \cite{Levy2002,Kim2011}, or thin-plate-splines \cite{Allen2003,Brown2007}. These approaches model the behavior of infinitesimally small portions of the object: For example, elasticity penalizes local stretch and bending, and thin-plate splines optimize for smooth deformations.

The problem with differential deformation models is that their assumptions are often not justified when considering shape families with substantial geometric variability, such as a diverse collection of four-legged animals (Figure~\ref{fig:teaser}). Elastic models can capture pose changes of a single shape reasonably well  \cite{Haehnel2003,Li2008}, but matching objects of different proportions creates strong artifacts (see Figure~\ref{fig:results-comparisons}d). Thin-plate splines \cite{Allen2003,Brown2007} are more flexible but their bias towards affine mappings still causes very noticeable artifacts (Figure~\ref{fig:results-comparisons}f). In both cases, reducing the weight of the regularization reduces bias but also increases noise and drift in the correspondences.

Isometry and conformal maps are by design already quite rigid: Both are already fixed by three point-to-point correspondences (for spherical topology), which is very valuable for solving the global matching problem efficiently \cite{Lipman2009}. These models are again useful for modeling pose changes, but shape sets of largely varying geometry are very unlikely to fall into the prescribed, low-dimensional sub-manifold of matchable shapes. Blending between partial maps~\cite{Kim2011} can reduce the problem, but substantial bias persists (see Figure~\ref{fig:results-comparisons}g).

Overall, computing dense correspondences among shapes of strongly varying geometry remains a problem that is mostly unsolved. The conceptual problem is that we need an effective notion of similarity that does not yet prescribe very specific geometric properties. Supervised machine learning from user annotated examples \cite{Kalogerakis2010,VanKaick2011,Sunkel2013} has shown promising results for establishing coarse correspondences. However, it cannot be easily extended to the dense case because it is very difficult if not impossible for a human to prescribe accurate dense correspondences for the training data.

This observation is a major motivation for our paper: We assume that coarse annotations are available. In addition to existing coarse matching methods \cite{Huang2011,Sidi2011}, we can always resort to manual human labeling. However, this is impossible for dense matches. We therefore develop a new method to get high-quality dense correspondences from a sparse and inaccurate initialization.

Towards this end we build upon another recent idea: Correspondence extraction from \emph{shape collections}. By considering many shapes of a similar kind simultaneously, more information is available. Several recent papers employ the \emph{cycle consistency} constraint to build correspondences in shape collections \cite{Nguyen2011,Huang2012,Kim2012}:
Correspondences are usually understood as a point-wise equivalence relation, being transitive over multiple shapes. Thus, unclosed loops indicate errors in pairwise matches that can be detected and removed. As pairwise regularizer, near-isometry \cite{Nguyen2011,Huang2012} or (optionally) extrinsic shape similarity \cite{Kim2012} are employed. However, this implicitly assumes that the shapes in the collection are dense samples of a continuous manifold of shapes, i.e., nearby samples are intrinsically very similar. This is not always the case in practical shape sets and thus introduces, as we will demonstrate experimentally, substantial artifacts.

In this paper, we therefore improve this model by explicitly regarding correspondence estimation as \emph{optimization of shape spaces}, aiming at capturing the class of observed models well. This can be understood as a statistical learning problem: A good explanation for a phenomenon is one that not only fits the observed data tightly but that is also simple \cite{Duda2000}. It is trivial to fit a large number of observations with a highly flexible model with lots of parameters (overfitting). However, making accurate predictions with a small and concise model makes such a hypothesis statistically meaningful.

Matching shapes of widely varying geometry forces us to choose mappings from a very large and sufficiently flexible set. However, from this large set, we aim at picking the simplest, the most \emph{compact representation}: The model should minimize the degrees of freedom utilized for representing the various shapes, rendering accidental matching unlikely: only natural correspondences will create simple shape spaces because they arise from a hidden, simple explanation for the observed geometric variability. Technically, this is formalized by minimizing the description length (MDL) of objects created by a Gaussian generative probabilistic model on a linear shape space. This approach has been originally developed in computer vision and medical imaging \cite{Kotcheff1998,Davies2002}.

In order to extend the applicability to a spectrum of typical computer graphics problems, we extend the original idea: First, we adapt the representation to handle meshes of generic topology.
Second, we introduce a part-based representation that permits modeling correspondences across shapes of \emph{varying} topology, interpreting each shape as an assembly of dockable, deformable parts. This allows us to learn a larger class of such composite models with both continuous (part deformation) and discrete (part assembly) variations. In particular, we introduce a novel algorithm to synthesize seamless and continuous models for assemblies of parts. Finally, the part-based approach yields high quality results: It decouples correlations between distant parts, which permits learning of expressive shape spaces with fewer examples, and with higher-quality correspondences.

In summary, we make the following main contributions: First, we introduce compact shape spaces for correspondence optimization to graphics, and demonstrate that this approach has a substantial impact on correspondence quality. Second, in order to make the method applicable to general meshes, we develop a new algorithm that can handle manifold meshes of generic topology while still maintaining meshing quality (uniform sampling and avoiding fold-overs). Third, we introduce a part-based formulation that represents shapes of variable topology; in particular, we describe new analysis and synthesis algorithms for composite shapes. We show that the part-based approach also improves the quality of the results over global optimization.
Finally, as an example application, we demonstrate an interactive system for designing deformable shapes with continuous and discrete variability.

\section{Related Work}

In this section, we discuss previous work on compactness of shape spaces, complementary to generic correspondence estimation methods already discussed above. The concept originates from studying point distribution models such as active shape/appearance models \cite{Cootes1995} that build generative Gaussian models of variability in images.

For model optimization, Hill et al.~\shortcite{Hill1994} have proposed compactness as criterion, and modeled this as the total variance of the shape distribution. Kotcheff and Taylor~\shortcite{Kotcheff1998} employ normal-distribution entropy, which creates sparse representations.

Davies et al.~\shortcite{Davies2002} refine this model by formulating the objective as minimum description length (MDL) approach \cite{Rissanen1978} that avoids inconsistencies and singularities.

Ericsson et al.~\shortcite{Ericsson2003} derive a gradient for the MDL energy, replacing the rather slow genetic algorithms and simplex methods by  more efficient gradient descent \cite{Heimann2005}.

The approach can be combined with surface parametrization~\cite{Davies2002b,Heimann2005,Davies2010} to handle manifolds and guarantee bijectivity, however this restricts the topology to the spherical case.

Cates et al. \cite{Cates2006} extend the approach to regularly sampled point-based representations of manifolds, handling the sampling uniformity by an elegant complementary entropy term. This approach also removes the topological restrictions but does not yield continuous, bijective mappings between meshes.

Our technique builds upon the entropy-based approach of Kotcheff and Taylor~\shortcite{Kotcheff1998}. Unlike previous methods, we use a smooth implicit representation of input meshes and parametrize the correspondences over a single such a shape of general topology. We enforce regular and uniform meshing by a bi-Laplacian regularizer and dynamic resampling \cite{Botsch2004}. Our representation automatically ensures cycle-consistent correspondences and permits handling of meshes of general topology while maintaining meshing quality (in practice, also effectively avoiding fold-overs). Further, the smoothness of the representation allows us to employ an efficient quasi-Newton method for optimization.

A problem of straightforward Gaussian MDL models is that they create bias towards linear representation of global shape rather than aligning surface features. Thodberg et al.~\shortcite{Thodberg2003} address this by adding a curvature-matching error. Our part-based approach can be seen as an alternative and complementary measure to limit such artifacts by providing localized adaptivity. In addition, it permits more flexibility in analyzing and representing composite shapes, which none of the previous methods provide.

A second, orthogonal problem is the global nature of the statistics. The model tends to overfit correlations between unrelated parts. For example, the poses of the arms in a human model are mostly independent, but excessive training data is required for a PCA model to recognize this. For this reason, many approaches have used part-based formulations \cite{Blanz1999,Zhang2004,Feng2008,Tena2011}. Our main contribution in this respect is that our analysis algorithm optimizes such models automatically. As a convenient by-product of the part-based correspondence optimization, our method optimizes the boundaries of the segmentation automatically given only a very coarse initialization. Further, our synthesis method works in the gradient-domain and thereby provides improved smoothness across boundaries in comparison to previous spatial domain methods \cite{Tena2011}.

\section{Creating Compact Shape Spaces}
\label{sec:basicMethod}

In this section, we describe the basic method for optimizing shape correspondences with the objective of creating compact shape spaces. We here first discuss the case of each shape consisting of a single part only; composite, part-based shape spaces will be discussed later, in Section~\ref{sec:partBased}.

\textbf{Input:} In the following, let $\mathcal{S}_1,...,\mathcal{S}_n \subset \reals^3$ be a set of 3D shapes. We assume that these are smooth, compact 2-manifolds embedded in $\reals^3$. The topology can be arbitrary but has to be fixed across all input shapes for now (by assembling multiple such parts, this requirement can be relaxed later).
In practice, the shapes are discretized as triangle meshes.
We denote the corresponding vertices by $\bv{S}_1,...,\bv{S}_n$; each $\bv{S}_i = (\bv{s}_1,...,\bv{s}_{n_i})$ is a matrix formed by the vector of the individual vertices. The set of triangles of each mesh are denoted by $\mathcal{T}_{\bv{S}_i}$.

\subsection{Linear Shape Spaces}

We first recap Gaussian generative shape models, as well known from literature \cite{Cootes1995,Blanz1999,Allen2003}, and define our notation.

First, we define the generative process: Let $\urshape$ be a \emph{urshape}, i.e., a base shape that has the same topology as each of the input shapes and that serves as parametrization domain for the shape space. This space is formed by the mappings:

\begin{equation}
  \shapeFun :  \urshape \times \reals^\sspaceDim \rightarrow \reals^3 \mbox{.}
  \label{eq:shapeFun}
\end{equation}

For each vector $\shapeParams = (\shapeParam_1,...\shapeParam_\sspaceDim) \in \reals^\sspaceDim$ and each $\vecx \in \urshape$, the function $\shapeFun$ returns a point on the generated shape. We assume the generative process to be linear. This shapes can be described by coordinates in an orthogonal basis. For a $\vecx \in \urshape, \shapeParams \in \reals^\sspaceDim$, we have:

\begin{equation}
  \shapeFunArgs{\vecx}{\shapeParams} = \shapeFunArgs{\vecx}{\shapeParam_1,...,\shapeParam_k} =  \meanShape(\vecx) + \sum_{i=1}^{d} \shapeParam_i \cdot \basisShape{i}(\vecx)
  \label{eq:subspace}
\end{equation}

Where the function $\meanShape$ encodes the mean shape and $\basisShape{1},...,\basisShape{\sspaceDim}$ are orthogonal basis functions that describe the possible linear modes of variation. In our implementation, we use (as most others) the mean shape as urshape, i.e., $\meanShape \equiv id$.
In practice, $\urshape$ will be approximated by a triangle mesh of $\numVertices$ vertices. We denote the $3 \times \numVertices$ matrix of the $n$ vertices of the mesh by $\urmesh$, and denote the created meshes by $\shapeFunArgs{\urmesh}{\shapeParams}$, and the continuous version by $\shapeFunArgs{\urshape}{\shapeParams}$, respectively.

We equip the shape space $\{\shapeFunArgs{\urshape}{\shapeParams} | \shapeParams \in \reals^\sspaceDim \}$ with a Gaussian probability measure with an axis aligned neg-log likelihood

\begin{equation}
  -\log \prob(\shapeParams) = \frac{1}{2}{\sum_{i=1}^{d} \frac{\shapeParam_i^2}{\stdDev_i^2}} + \text{const}\mbox{,}
  \label{eq:loglikelihood}
\end{equation}

where $\stdDevs = (\stdDev_1,...,\stdDev_\sspaceDim)$ specifies the standard deviations along the main axes of the model.

Further, we will usually consider the space of shapes generated by $\shapeFun$ and then rigidly arranged in $\reals^3$. Given $\rot \in O(3)$ and $\vect \in \reals^3$, we denote a rigidly transformed shape (in slight abuse of notation) by:

\begin{equation}
  \rot\left(\shapeFunArgs{\urshape}{\shapeParams}\right) + \vect := \left\{ \rot \cdot \shapeFunArgs{\vecx}{\shapeParams} + \vect \ \mid \ \vecx \in \mathcal{U} \right\}\mbox{.}
  \label{eq:transfShape}
\end{equation}

\textbf{Building the model:} Given a set of input shapes and correspondences between them, we can easily build Gaussian shape spaces using principal component analysis (PCA): Assume that we are given a set of consistently triangulated vertex meshes $\bv{S}_1^*,...,\bv{S}_n^*$ that match the input $\mathcal{S}_1,...,\mathcal{S}_n$ with vertex correspondence, i.e., corresponding vertices located at matching geometry (we use the star to denote known correspondences). We compute the mean $\overline{\bv{S}}$ by averaging the input shapes and determine the covariance matrix

\begin{equation}
  \Sigma_\bv{S} = \frac{1}{n-1}\sum_{i=1}^{n} \left(\bv{S}_i^* - \overline{\bv{S}}\right) \left(\bv{S}_i^* - \overline{\bv{S}}\right)^{\operatorname{T}}
  \label{eq:pcaCov}
\end{equation}

The mean and the eigenvectors of $\Sigma_\bv{S}$ yield the mean and basis meshes and the eigenvalues correspond to the standard deviations $\stdDevs$. Further, it is easy to see (for example, by applying a singular value decomposition and rearranging terms) that the Gram matrix

\begin{equation}
  \bv{G}_\bv{S} = \sum_{i=1}^{n} \left(\bv{S}_i^* - \overline{\bv{S}}\right)^{\operatorname{T}} \left(\bv{S}_i^* - \overline{\bv{S}}\right)
  \label{eq:pcaGram}
\end{equation}

has the same eigenvalue spectrum (up to the factor $n-1$). In the continuous case, the sum is replaced by an integral. Assume that we have homeomorphisms $s_i^*: \urshape \rightarrow \mathcal{S}_i$ that encode continuous correspondences to our input shapes $\mathcal{S}_1,...,\mathcal{S}_n$. We then again form the mean function $\overline{s}$ by averaging and the $n \times n$ Gram matrix:

\begin{equation}
  \bv{G}_\mathcal{S} = \int_{\urshape} \left(s_i^*(\vecx) - \overline{s}(\vecx)\right)^{\operatorname{T}} \left(s_i^*(\vecx) - \overline{s}(\vecx)\right) d\vecx \mbox{.}
  \label{eq:pcaGramCont}
\end{equation}

The matrix has at most rank $n$; in the (typical) case of redundancy in the shape collection, the number of significant eigenvalues will typically be substantially smaller than $n$. Importantly, the spectrum does not just depend on the geometry of $\mathcal{S}_1,...,\mathcal{S}_n$ but crucially on the correspondences encoded in the functions $s_1^*,...,s_n^*$. While the variability of the shapes prescribes a lower bound on the rank of $\bv{G}_\mathcal{S}$, we can in general artificially inflate it up to full rank by just letting the correspondences drift randomly along the surface.

\subsection{Compactness}
\label{sec:Compactness}

We now discuss how to measure the compactness of the shape space and how to minimize it. We also recap ideas from \cite{Kotcheff1998,Cates2006,Davies2010} to keep the paper self-contained.

\textbf{Spectral view:} Let $\stdDevs$ denote the vector of eigenvalues of $\bv{G}_\mathcal{S}$. If the correspondences $s_1^*,...,s_n^*$ include unnecessary movements along the surfaces of the objects, the spectrum will spread out, creating more non-zero eigenvalues. In reverse, a compact shape space should have a compact spectrum. A simple way of modeling this is to penalize the square norm $\| \stdDevs \|^2_2 = \operatorname{tr}(\bv{G}_\mathcal{S})$ of the eigenvalues, as proposed by Hill and Taylor~\shortcite{Hill1994}. It is equivalent to trying to keep all surface points in deformed shapes close to the mean shape, independent of each other (therefore not transporting information globally). From a spectral perspective, it favors multiple small eigenvalues over a few large ones, which does not match the intuition of a low-dimensional generative process that we want to reconstruct. Rather than that, we aim at a sparse spectrum, as detailed next.

\begin{figure}
  \centering
  \begin{tabular}{ccc}
    \myincludegraphics[width=0.3\linewidth]
    {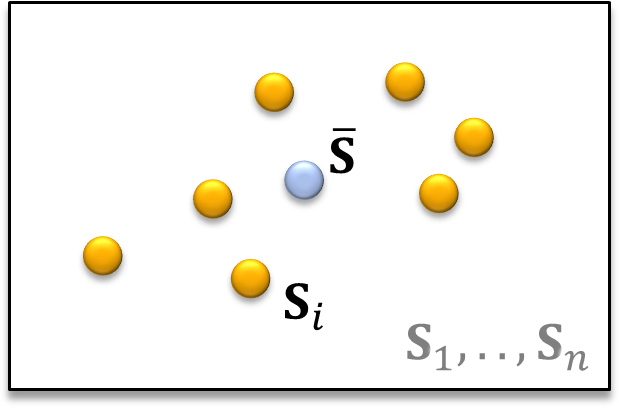} &
    \myincludegraphics[width=0.3\linewidth]
    {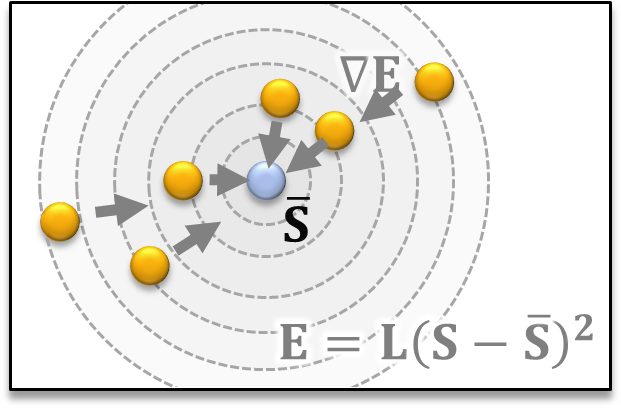} &
    \myincludegraphics[width=0.3\linewidth]
    {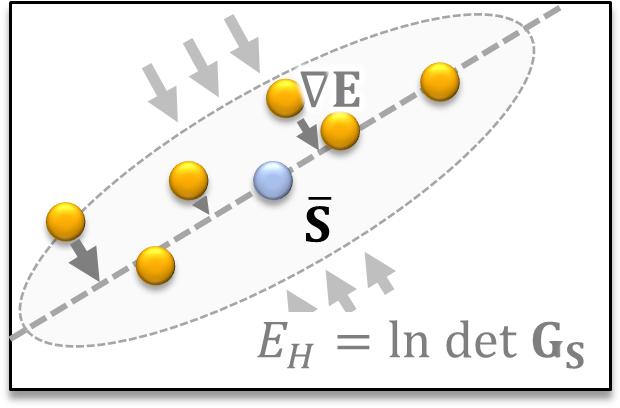} \\
    \small (a) untransformed &
    \small (b) linear differential prior &
    \small (c) entropy prior
  \end{tabular}
  \caption[Geometric interpretation of the entropy prior]{Geometric interpretation: Differential models such as (linearized) elasticity or thin-plate splines impose a quadratic energy on a linearly transformed shape space that attracts all target shapes to the source shape. Our method minimizes the entropy of the ensemble, moving correspondences such that the shapes align in a lower-dimensional subspace, creating less bias.}
  \label{fig:shapeSpace}
\end{figure}

\textbf{Probabilistic view:} We can also look at the probability distribution the shapes are drawn from. The less variability it permits, without reducing the likelihood of the training examples, the more concisely it captures the shape space. In this view, we should measure the \emph{entropy} of the Gaussian model:

\begin{align}
  H_{\prob(\shapeParams)} &
  = \frac{1}{2} \ln  \prod_{i=1}^{n} \stdDev_i^2 + const.
  \label{eq:entropy}
\end{align}

This approach suffers from singularities: If one of the eigenvalues becomes zero, the determinants of the covariance and Gram matrix become zero, leaving the entropy ill-defined. Further, driving even just the least eigenvalue close to zero would falsely indicate a near-perfect solutions, which leads to instability and inconsistency.

\textbf{Information theoretic view:} From the point of view of information theory, we can measure the capacity of the generative probabilistic model (Equations~(\ref{eq:subspace},~\ref{eq:loglikelihood})) to encode different models by considering the description length of a specific shape, given the knowledge of the generative model in terms of the probabilistic shape space. To transmit one shape, we need to encode the shape parameters $\shapeParams=(\shapeParam_1,...,\shapeParam_\sspaceDim)$. Given an independent Gaussian distribution along each axis $\basisShape{i}$ with variance $\stdDev_i^2$, and assuming that a finite accuracy of $\Delta > 0$ is required in our application, encoding a single parameter requires roughly $\mathcal{O}(\log \frac{\stdDev_i}{\Delta})$ bits \cite{Thodberg2003b} (see Davies et al.~\shortcite{Davies2002} for the accurate and more detailed derivation). For small variances $\stdDev_i < \Delta$, no information needs to be encoded. This suggests the following energy \cite{Kotcheff1998,Cates2006} that approximates the information content of the shape space depending on correspondences $s_1,...,s_n$:

\begin{align}
  E_{H} (s_1,...,s_n) &= \ln \prod_{i=1}^{n} (\stdDev_i + \delta)    = \ln \det \left( \bv{G}_\mathcal{S} + \delta \bv{I} \right)
  \label{eq:entropyEnergy}
\end{align}

$\delta > 0$ is a regularizer that determines the accuracy of the shape space: We assume that independent of the example data, there is always an isotropic Gaussian noise component of standard deviation $\delta$ in all dimensions of the space. This removes the singularity and makes the entropy usable as measure that encourages sparse PCA spectra during correspondence optimization \cite{Kotcheff1998}. This is an approximation to coding length \cite{Davies2002}; nonetheless, it already yields favorable results in practice.

\textbf{Geometric view:} We can also interpret these results as imposing a prior in a shape space. Figure~\ref{fig:shapeSpace} shows schematically a number of example shapes $\bv{S}_1,...,\bv{S}_n$ as points in a high-dimensional shape space. Traditional regularizers such as thin-plate-splines or linearized elasticity impose a Gaussian prior, i.e., the neg-log-likelihood is a quadratic energy of the form

\begin{equation}
  E(\bv{S}_1,...,\bv{S}_n) = \sum_{i=1}^n\left(\bv{L}(\bv{S}_i - \overline{\bv{S}})\right)^2 \text{,}
  \label{eq:}
\end{equation}

where $\bv{L}$ is a linear operator (a matrix) that acts on the vertex sets $\bv{S}_i$ interpreted as $(3\cdot n_i)$-vectors. For example, in thin-plate-splines, $\bv{L}$ measures the bending by taking second derivatives. In other words, traditional (linear) differential priors can be seen as an isotropic attraction to a single point (the urshape) in a linearly transformed shape space (Figure~\ref{fig:shapeSpace}b). Contrarily, minimizing the entropy encourages a tight fit of an ellipsoid to the data, minimizing its volume, and thereby encouraging all models to be located on a low-dimensional linear subspace (Figure~\ref{fig:shapeSpace}c). This creates bias towards a linearly correlated representation rather than towards a single shape. It is not surprising that this yields significantly better results when the final objective is to describe a shape collection with exactly this representation rather than reconstructing it from pairs of biased, point-wise matches in shape space.

{
  \begin{figure}
    \centering
    \small
    \renewcommand{\tabcolsep}{1.4em}    \begin{tabular}{ccc}
      \myincludegraphics[scale=1.3]
      {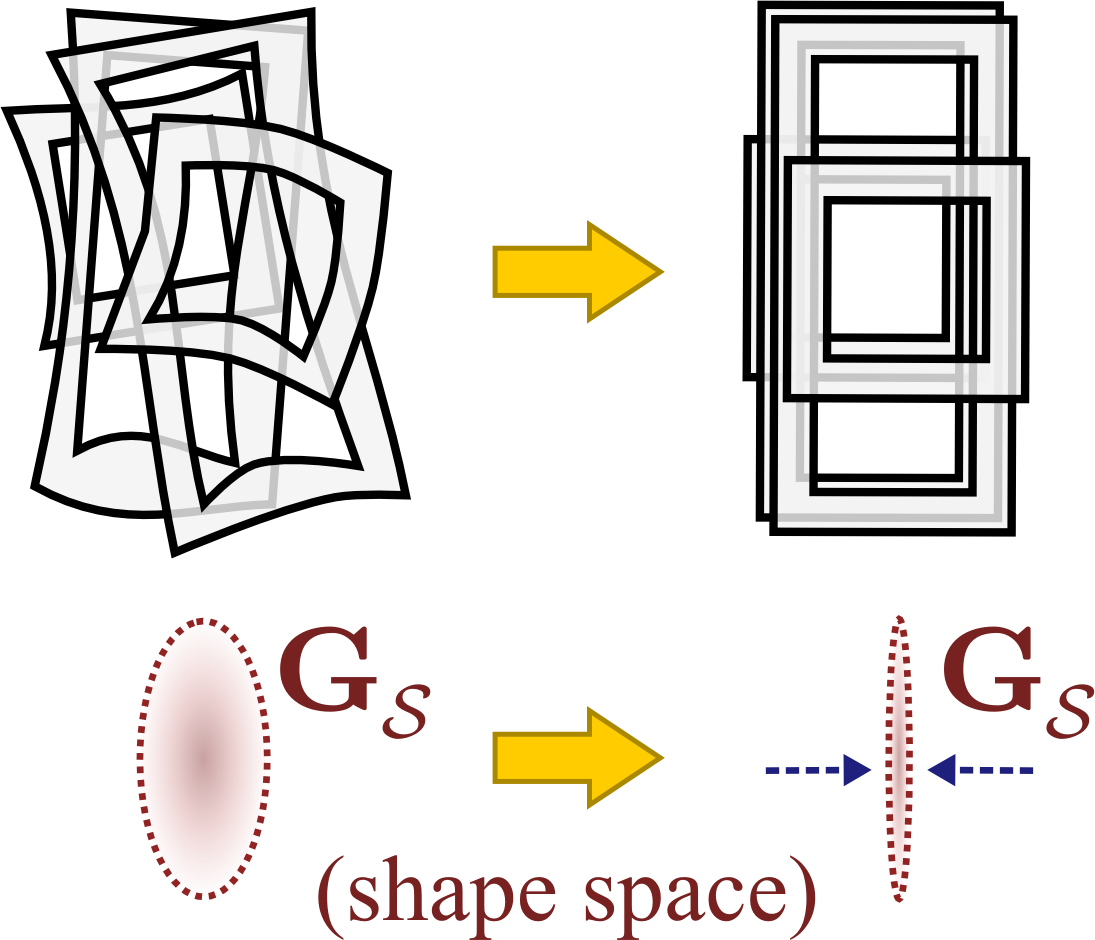} &
      \myincludegraphics[scale=1.3]
      {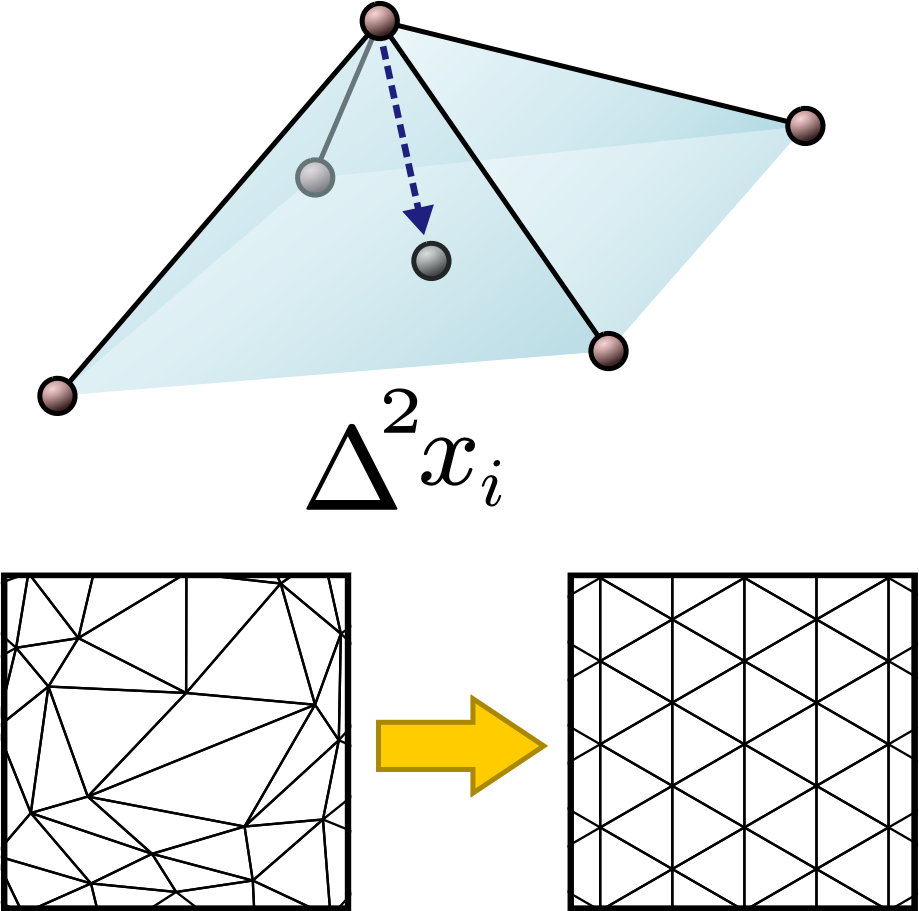} &
      \myincludegraphics[scale=1.3]
      {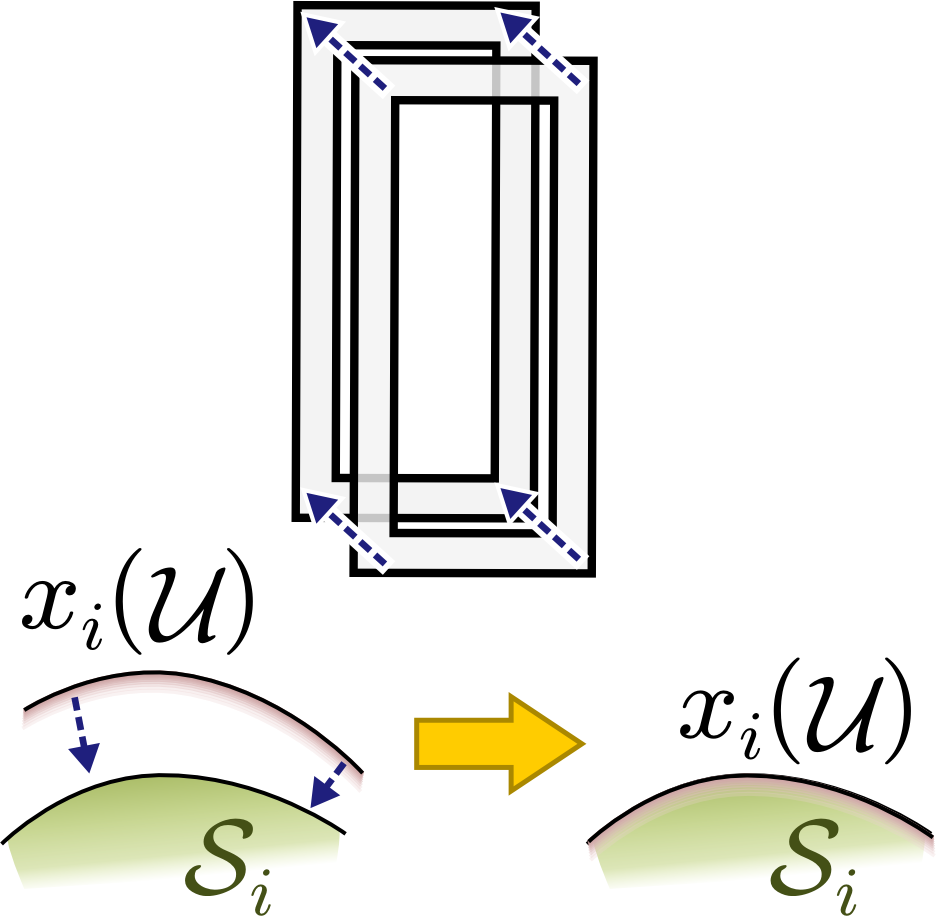} \\
      (a) $E_H(\allCorr)$ &
      (b) $E_L(\allCorr)$ &
      (c) $\corr_i \in \mathcal{S}_i$
    \end{tabular}
    \caption[Shape space optimization energy terms]{Shape space optimization. Our energy consists of three terms (a-c) and a hard constraint (d). (a) The term $E_H$ minimizes the description length of the model by aiming at minimizing the entropy (volume) of the Gram matrix of the model collection. (b) $E_L$ encourages uniform meshing using a Laplacian energy.
      (c) Matching the data surfaces $\mathcal{S}_i$ is a hard constraint.}
    \label{fig:opt_EH}
  \end{figure}
}

\subsection{Shape Optimization}
\label{sec:ShapeOptimization}

Let $\bv{S}_1,...,\bv{S}_n$ be a set of example models given as triangle meshes. We approximate these by smooth surfaces $\mathcal{S}_1,...,\mathcal{S}_n$, as detailed later. Let $\bv{U}$ be an urshape of matching topology. We denote the vertices of $\bv{U}$  by $\bv{u}_1,...,\bv{u}_m$. We now want to compute correspondences

\begin{equation}
  \corr_i\colon \{\bv{u}_1,...,\bv{u}_m\} \rightarrow \mathcal{S}_i \mbox{.}
  \label{eq:corrFun}
\end{equation}

We denote the set of all correspondences by $\allCorr = \{ \corr_1,..,\corr_n \}$. All of these are  hard-constrained to be located on the (smoothed) input surfaces. We optimize the correspondences by minimizing the following energy, subject to the constraint of moving only along the surface (as illustrated in Figure~\ref{fig:opt_EH}):

\begin{equation}
  E(\allCorr) = E_H(\allCorr) + \mu_L E_L(\allCorr)
  \label{eq:energy}
\end{equation}

The term $E_H$ approxiates the description length as discussed above and $E_L$ is a bi-Laplacian regularizer.

We set its weight $\mu_L$ to the ratio of the number of triangles divided by the surface area squared (to make the overall weight mesh-independent), multiplied by a relative weight of $0.25\cdot 10^{-5}$.

The overall energy is minimized using l-BFGS, a nonlinear quasi-Newton solver. Further, we factor out rigid motions according to Equation~(\ref{eq:transfShape}): We compute a least-squares optimal translation, rotation, and reflection from the initial correspondences. The rigid motion is updated during the optimization by including the rotation as variable in the optimization (parametrization the small rotational update as Euler angles with respect to the initial least-squares fit).

\subsubsection{Compactness}

For creating compact shape spaces, we use the energy $E_H$ from Equation~(\ref{eq:entropyEnergy}). We compute the Gram matrix by integrating over the deformed triangle meshes according to Equation~(\ref{eq:pcaGramCont}). Because of additional regularization (described next), it is sufficient to approximate the integrals by an unweighted sum over vertex positions (Equation~(\ref{eq:pcaGram})). We compute the derivative of the energy using the explicit formula derived in \cite{Kotcheff1998}.

\subsubsection{Regularization}

The regularization term $E_L$ is a prior on the graph Laplacian of the deformed meshes $\corr_i(\urmesh)$. With $N_i$ denoting the set of indices of vertices sharing an edge with vertex $\bv{u}_i$ in the mesh $\urmesh$, we obtain:

\begin{equation}
  E_L(\allCorr) = \sum_{i=1}^{n} \sum_{j=1}^{m} \frac{1}{|N_j|}\left( \sum_{k \in N_j} \big(\corr_i(\bv{u}_j)-\corr_i(\bv{u}_k)\big) \right)^2
  \label{eq:laplaceReg}
\end{equation}

This term encourages the graph Laplacian of the triangle mesh to be zero, which is the case if every vertex is located in the center of its 1-ring neighborhood, corresponding to a uniform triangulation \cite{Botsch2004}. Although adding this least-squares energy does not guarantee bijectivity of the mapping, it also effectively avoids fold-overs in practice.

\subsubsection{Data Modeling}
\label{sec:dataModeling}

We model the hard-constraint that correspondences must remain on the input surfaces by a level-set approach. As our input is only discrete, $C^0$ mesh approximation of a shape, we first build a smooth surface $\mathcal{S}_i$ that tightly approximates $\bv{S}_i$ so that we can slide along the surface smoothly during optimization. We first sample the input mesh with a dense, uniform point cloud $\bv{S}_i'=\{ \bv{s}'_1,...,\bv{s}'_{n_i} \}$ representing the input mesh $\bv{S}_i$ with (given) oriented normals $\{ \bv{n}_1,...,\bv{n}_{n_i} \}$. We fit a signed distance function $d: \reals^3 \rightarrow \reals$ to $\bv{S}_i$ by minimizing the following energy \cite{Calakili2011}:

\begin{align}
  E(d) = \mu_z\underbrace{\sum_{i=1}^{n_i} d(\bv{s}'_i)^2}_{\text{zero crossing}} + \mu_g\underbrace{\sum_{i=1}^{n_i} \left \| \nabla d(\bv{s}'_i) - \bv{n}_i \right \|^2}_{\text{gradients}}
  \nonumber \\
  + \mu_F \underbrace{\int_\Omega \| H_d(\vecx) \|_F^2 d\vecx}_{\text{smoothness}}
  \label{eq:signedDistanceField}
\end{align}

The first term assures that the zero crossing of $d$ is at the data points. The second term aligns the gradients with the normals, creating a smooth result and removes the trivial solution ($d=0$). The last term integrates the squared Frobenius norm of the Hessian of $d$ over a bounding volume $\Omega$, acting at a regularizer that propagates function values linearly and encourages smoothness. We set the weight $\mu_z=1/n_i$, $\mu_g=0.1/n_i$ and $\mu_H= 10^{-4}/|\Omega|$, which is sufficient to smooth very sharp corners a bit.

We optimize this quadratic energy by solving the linear system resulting of a finite difference discretization with spacing $h$ set to $h=1.5\%$ of the bounding box of the object. Continuous values for $d(\vecx)$ and $\nabla d(\vecx)$ are obtained by interpolation with radial basis functions at each grid point; we employ Wendland kernels $\max(0,(1-\|\vecy - \vecx\|^2/h^2)^3)$. $H_d$ is approximated by finite differences over the grid. The domain $\Omega$ is obtained by including all grid cells within a distance of $4h$ to data points. Triangle meshes sampled with spacing $h/4$ to obtain $\bv{S}_i'$. We refer to the zero-level set of the result as $\mathcal{S}_i$.

\textbf{Using the implicit function:} The $\mathcal{S}_i$ serve as constraint manifolds for correspondences during optimization: First, any initial solution is projected to $\mathcal{S}_i$ by simple gradient descent. During numerical optimization, the quasi-Newton l-BFGS solver attempts to update the correspondences positions: $x_i(\bv{u}_j) \rightarrow x'_i(\bv{u}_j)$ by first finding a new direction and then the distance by a line search.
In each iteration, we project $x'_i(\bv{u}_j)$ back onto the surface (using the exponential map in $x_i$). The rational is that small step sizes turn the smooth constraint into a sequence of linear subspace constraints that can are handled by the quadratic (low-rank) optimizations performed in the inner loop of l-BFGS.

\textbf{Motivation:} In experiments with various formulations, the implicit function formulation with hard constraints to the zero level-set turned out to be most reliable and crucial for good results. Other options did not give satisfactory results: Least-squares soft-constraints are unreliable: weak constraints have trouble with thin structures and sharp creases, and strongly weighted soft constraints yield a numerically ill-conditioned energy, preventing convergence. The option of just using the input triangle meshes was not satisfactory either: Using such a $C^0$ surface lead to spurious local optima in our experiments. Experiments with a projection to dynamically computed MLS-approximation of the surface have also turned out to be slow and unreliable for general surfaces with small feature size.

\section{Extended Model}
\label{sec:partBased}

We now extend our approach by introducing composite, \emph{part-based} models that capture correspondences among objects of varying topology.

\begin{figure}
  \small
  \centering
  \begin{tabular}{cccc}
    \myincludegraphics[height=4.5cm]
    {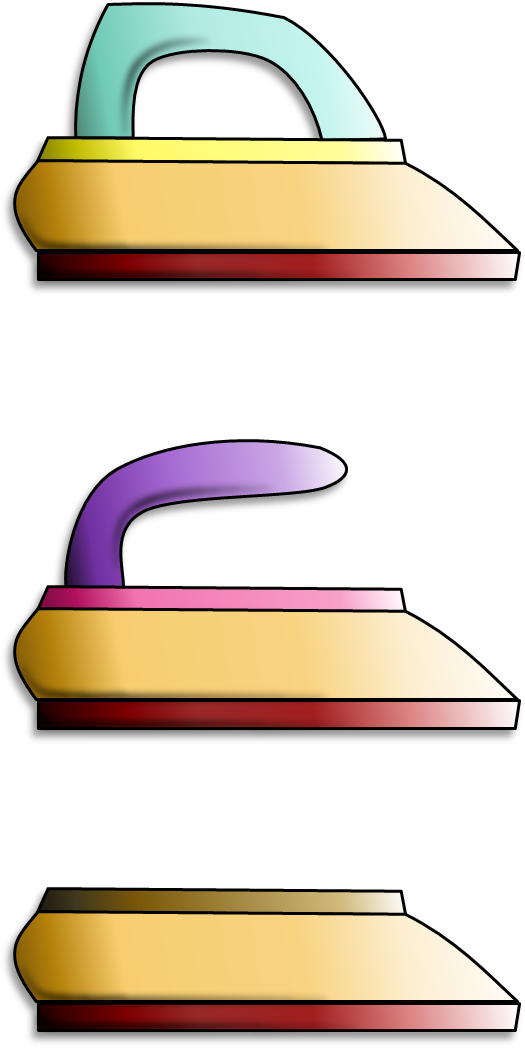}  &
    \myincludegraphics[height=4.5cm]
    {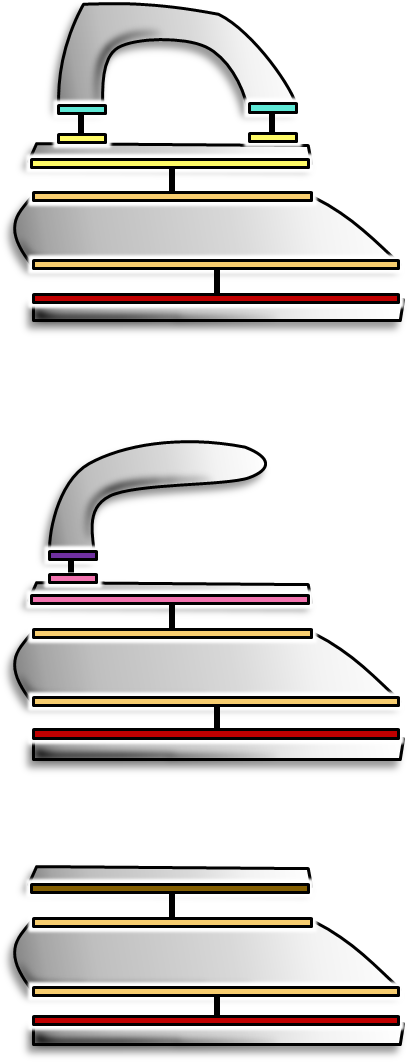}  &
    \myincludegraphics[height=4.5cm]
    {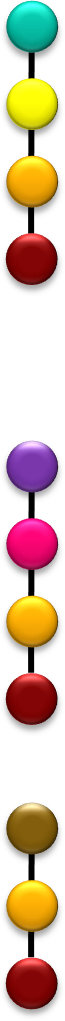}  &
    \myincludegraphics[height=4.5cm]
    {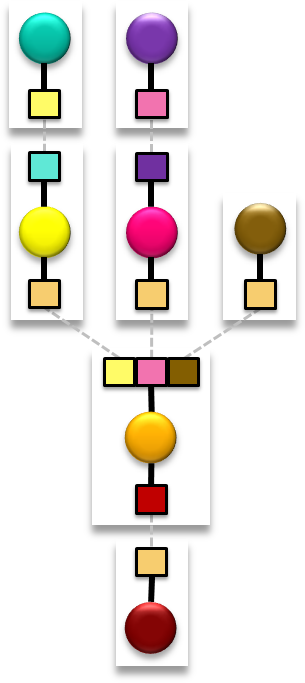} \\     (a) input shapes &
    (b) docking sites &
    (c) part graphs &
    (d) docking rules
  \end{tabular}
  \caption[Decomposing objects into morphable parts]{Decomposing objects into morphable parts (iron example; c.f. Figure~\ref{fig:results}b). (a)~Objects are segmented into different parts, indicated by different colors.(b)~Boundaries permit docking of shapes according to the observed variations. (c) Each discrete assembly corresponds to a graph of parts. (d)~This yields a set of rules for possible part arrangements.}
  \label{fig:parts}
\end{figure}

\subsection{Part-Based Modeling}
\label{sec:PartBasedModeling}

The method as discussed so far, as well as previous proposals in literature, is restricted to shapes that have global correspondences and form a single, global shape space. In practice, this is often a strong restriction. Many man made shapes consist of composite parts (for example, the irons in Figure~\ref{fig:results}b have been assembled from different parts), forming shape spaces of varying topology that cannot be captured by a single shape space.

We therefore propose a model that decomposes complex shapes into a set of \emph{parts} that have individual shape spaces. First, we modify the analysis algorithm to optimize both the shape and the decomposition of the surface. Second, develop a synthesis algorithm that can build seamless models consisting of deformed parts in different poses. The synthesis can handle general arbitrary constraints (changing the discrete composition of the parts, handles for free-form deformation, subspace constraints).

\subsubsection{Analysis}
\label{sec:Analysis}

\textbf{Input:} We again assume that we are given a set of example shapes $\bv{S}_1,...,\bv{S}_n$ as triangle meshes (Figure~\ref{fig:parts} shows an example, reflecting the actual result demonstrated in Figure~\ref{fig:results}b). We further assume that the shapes are segmented into parts, i.e., every triangle is tagged with a part type $p \in \{1..K\}$, where $K$ is the number of different part types (Figure~\ref{fig:parts}a shows the types as different colors). Each part of the same type must have the same topology (The irons example use a adapter pieces (yellow/pink/brown) to attach handles of different topology to the body). Each discrete configuration corresponds to a different graph of parts (Figure~\ref{fig:parts}c). In addition, each part has continuous parameters (not shown) that permit deformation according to the shape space learned from all parts of the same type (same color in our figures).

The initialization of the part boundaries does not need to be precise; only the topology and coarse geometry needs to match. We will improve the segmentation geometry automatically.

\textbf{Part docking:} Parts will share common boundaries, and possibly in different combinations. We learn the way parts can be discretely assembled from the input by just reading of the observed adjacency relations from the input.

Figure~\ref{fig:parts}d,e show the rules that have been deduced from the input. Boundaries between that connect two types of parts across a common docking site are always in fixed correspondence; i.e., the dense correspondences of the parts themselves are enforced at the boundary, too.

In the following we discuss our analysis algorithm that creates part graphs and shape spaces for each part automatically given a coarse user segmentation and possibly a few additional landmark matches.

\begin{figure}
  \centering
  \myincludegraphics[width=0.70\columnwidth]
  {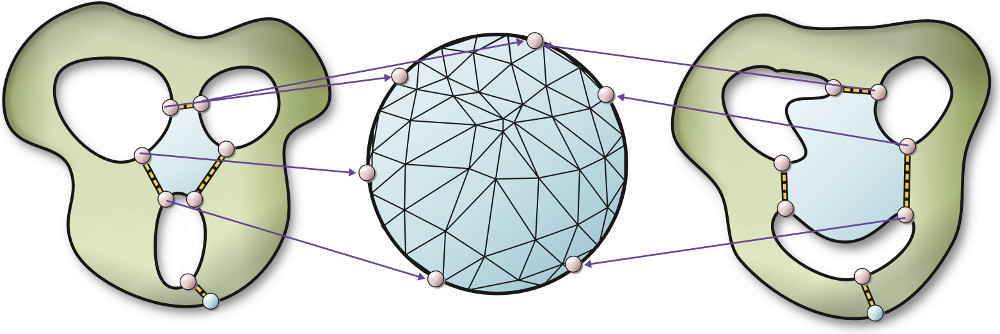}
  \caption[Cutting to disc topology]{In order to obtain initial correspondences, we cut the parts into pieces of disc topology. Afterwards, a cross parametrization ensures a bijective initialization.}
  \label{fig:param}
\end{figure}

\textbf{Part parametrization:} The first step is to compute initial dense correspondences. We need bijective correspondences without fold-overs. For this, we use cross-parametrization \cite{Kraevoy2004}: We first cut the parts further into topological discs, and then compute a cross-parametrization of the discs to obtain initial correspondences (see Figure~\ref{fig:param}).
For cutting, we first detect the interior boundaries within all parts. We connect each resulting boundary curve to its closest neighbor (see Figure~\ref{fig:param}) and then cut along a geodesic path between the corresponding closest point. Cutting is iterated until only topological discs remain, and the process is done in all parts simultaneously.
This initialization is presented to the user, who can move the initial landmark correspondences along the boundaries (red dots in Figure~\ref{fig:param}). The resulting sub-parts are set into correspondences by a least-squares conformal map to a unit circle; the boundaries of the circle are set into correspondence by comparing the relative arc length (normalized to $[0,2\pi]$), using the cutting points as starting point. For the outer boundary $\partial \partMesh^{(p)}_j$ of the initial segmentation, the user has to specify this starting point manually.

The result of this step are dense correspondences between all parts, with topology consistent to the user-defined segmentation. The correspondences are guaranteed to be bijective, but the quality is usually very bad, showing strong drift and distortions across the shape.

\textbf{Optimization:} We now perform the optimization from Section~\ref{sec:basicMethod} to improve this initial guess. For each part type (same color in our figures), we setup a separate energy $E^{(p)}$ according to Equation~(\ref{eq:energy}). We use the cross-parametrization result as initial correspondences $\corr_i^{(p)}$, and the resulting mean shape as urshape $\urmesh^{(p)}$. We first run the optimization separately, constraining the boundaries of the domain to fit the boundary curve by a point-to-line energy that snaps the closest vertex to its the boundary of the parts. We add the boundary energy
\begin{equation}
  E_B = \sum_{p=1}^{K} \sum_{j=1}^{n_p} \int_{\partial \urmesh^{(p)}} \hspace{-1mm} \operatorname{dist}\left(\corr_j^{(p)}(\vecx), \partial \partMesh^{(p)}_j\right) d\vecx
  \label{eq:boundaryS}
\end{equation}
that measures the deviation of boundary vertices from the boundaries of the input parts.

\textbf{Boundary optimization:} After the energy has converged, we remove the constraint of Equation~(\ref{eq:boundaryS}) and start optimizing the boundary location. We need to make sure that parts still meet at the boundaries, and this should happen in a consistent way. As consistency condition, we maintain fixed correspondences along all matching part boundaries of the same type. We impose this consistency as a soft constraint in an alternating two-stage optimization:

In stage one, we find all pairs of closest points between boundaries of matching type: For points $\vecx \in \boundarysym_l^{(p)}$ we compute all instances in the data, and for each instance, the closest point in the adjacent instance of type $\boundarysym_l^{(p)}$. We average over all of these matches and set a soft constraint penalizes the quadratic distances for all these pairs.
In stage two, we run the optimization of $E^{(p)}$ according to Equation~(\ref{eq:energy}) with the additional constraint energy added. We optain improved correspondences, which are again used to refine the correspondences.

Conceptually, this could be interpreted as a variant of iterative closest points (ICP), performed simultaneously along multiple boundary curves while keeping their correspondences consistent. The alternating estimation of boundary correspondences is combined with the estimation of global rigid motions for each part, as already introduced in Section~\ref{sec:ShapeOptimization}.

\textit{Further details:} In our implementation, a few extra steps are performed to improve the efficiency of the method. First, before computing the cross-parametri\-zation, we use the dynamic remeshing algorithm of Botsch et al.~\cite{Botsch2004} in order to improve the mesh quality of each part (which might already have been bad in the input meshes). The method iteratively minimizes the sum of the squared graph Laplacians and performs edge contractions / vertex splits in order to create a uniformely sampled mesh. We use the average edge length in all part instances as length criterion. Second, after parametrization, we might end up with very uneven sampling; the conformal map can have large scale factors that lead to a uneven distribution of triangles. Using vertex splits, again according to the same criterion, we refine regions that are undersampled and project the resulting newly inserted points onto the implicit surface that models the data. These steps could in principle be omitted, but then a very dense initial mesh is required to obtain results of good quality.
Another improvement is to perform a final optimization pass of the correspondence energy $E(\allCorr)$ (Equation~\ref{eq:energy}) where the Laplacian regularizer is evaluated for each composite input shape $\bv{S}_i$ instead of its parts separately, which makes sure that the boundaries between parts are only determined by the compactness criterion and not by the mesh regularization (we obtain a slight improvement here).

Overall, the result of the preceding is an optimized composite shape space in which (i) the correspondences within each part, (ii) the position of the boundaries on the example shapes, and (iii) the correspondences among matching boundaries have been optimized with the goal of compactness and mesh quality. In the following, we discuss how we can utilize the result to create new shapes.

\subsubsection{Synthesis}
\label{sec:Synthesis}

The model that we have obtained in the previous step describes a shape by a set of parts that are connected along their boundary lines. A key feature of this extended model is that we can instantiate composite models consisting of multiple parts, potentially rearranged by attaching the parts differently across compatible boundaries. We therefore need to devise a generative process by which we can instantiate such composite shapes, governed by multiple local shape spaces. We aim at maximum flexibility: Given an arbitrary arrangement of parts and arbitrary user constraints on geometry and shape parameters for each part, we want to find a global geometry that fits all of these constraints best.

\subsubsection*{Variational Part Reconstruction}

The first step is to formulate the problem of reconstructing the part shapes as a variational problem. In order to facilitate a smooth reconstruction later on, we formulate the whole process in the gradient domain~\cite{Sumner2005,Sorkine2007}. We first consider a single part. Assume that we are given a part shape space by its urshape $\urmesh$, its mean and variation modes $\meanShape,...,\basisShape{\sspaceDim}$, and the standard deviations $\stdDevs=(\stdDev_1,...,\stdDev_\sspaceDim)$. Our objective is to reconstruct an instance $\bv{V} = (\vecv_1,...,\vecv_n)$ of this shape with shape parameters $\shapeParams = (\shapeParam_1,...,\shapeParam_\sspaceDim)$. Because of the formulation as an optimization problem, multiple parts can be coupled along boundaries, thereby implicitly constraining the reconstruction and finding the best embedding of the part graph in a least-squares sense.

We model the similarity by comparing each vertex $\vecv_i$ with a reconstructed vertex $\meanShape(\vecu_i) + \sum_{k=1}^{\sspaceDim} \shapeParam_k \basisShape{k}(\vecu_i)$. The residual is minimized in a least squares sense. In order to get a smooth transition between multiple parts later, we follow \cite{Sumner2005} and formulate the optimization in a gradient domain. We do not compare absolute coordinates but edge vectors in the mesh. Finally, we also include an orthogonal transformation $\bv{R}$ to be invariant to rigid motion (translational invariance is automatically obtained by working on edge differences).

Formally, we get the following energy:

{\small
  \begin{align}
    E_{rec}(\bv{V}, \shapeParams, \bv{R}) = \sum_{i=1}^{n} \sum_{j \in N_i}
    \omega_{i,j} \Bigg( \big( \vecv_i - \vecv_j \big)
    -  \bv{R} \big(\meanShape(\bv{u}_i) - \meanShape(\bv{u}_j) \big) \nonumber \\
    -\sum_{k=1}^{d} \shapeParam_k \bv{R} \left(\basisShape{k}(\vecu_i) - \basisShape{k}(\vecu_j) \right)  \Bigg)^2 + \sum_{i=1}^{\sspaceDim} \frac{\lambda_i^2}{2\stdDev^2}
    \label{eq:rec}
  \end{align}
}

Again, $N_i$ denotes the set of neighboring vertices of $\bv{u}_i$ in the mesh $\urmesh$. $\omega_{i,j}$ is the cotangent weight of the edge $(\bv{u}_i,\bv{u}_j)$ in the mean shape. The variable $\bv{R}$ is a rotation variable, to be optimized along with the variables $\bv{V}$ and $\shapeParams$.

This formulation is an extension of the as-rigid-as-possible shape deformation of \cite{Sorkine2007}, encouraging the result to be as-close-as-possible to a linear subspace of models (ignoring rigid differences as well). We use the same optimization method: The linear system is solved alternatingly with an update of the rotation matrix $\bv{R}$ (which, in our approach, is global for the whole part); see Sorkine and Alexa's paper \shortcite{Sorkine2007} for details.

\begin{figure}
  \centering
  \begin{tabular}{ccc}
    \myincludegraphics[scale=1.4,valign=B]
    {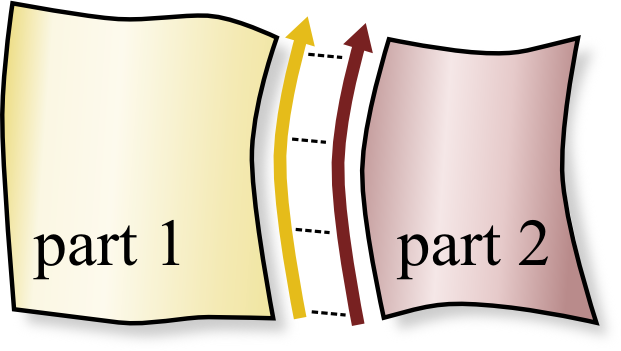} &
    \myincludegraphics[scale=1.4,valign=B]
    {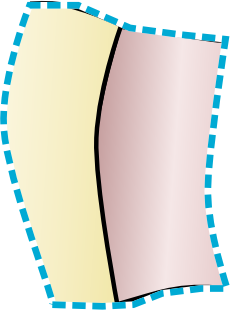} &
    \myincludegraphics[scale=1.4,valign=B]
    {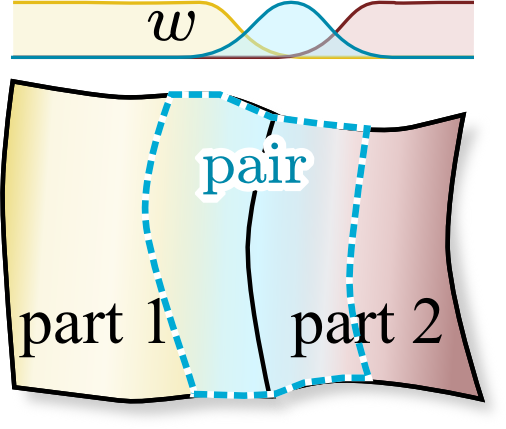} \\
    (a) mesh    & (b) pair  & (c) docking  \\
    boundaries & geometry & with weights
  \end{tabular}
  \caption[Attaching two part instances]{Attaching two part instances: (a) Boundaries are joined continuously by sharing variables. (b) We extract geometry adjacent to the boundary and (c) blend between part and pair shape spaces for smoother transitions.}
  \label{fig:dockingSites}
\end{figure}

\subsubsection*{Reconstructing Part Graphs}

In order to reconstruct shapes consisting of multiple parts, we add up the energies $E_{rec}$ for all parts. Along the boundaries, the analysis stage gives us fixed and consistent correspondences. Therefore, we can remesh the urshapes of the parts such that they share common points along the boundaries. We then use the same variables in order to enforce a $C^0$ continuous solution (see Figure~\ref{fig:dockingSites}a).

\textbf{Improved smoothness:} Although the shape of the boundary curve transports information between pairs of parts, it only captures limited information on the correlation between the part shapes. We therefore learn a more expressive model from the input data: We form an extended region by gathering the geometry within a fixed distance to the boundary between the pair of parts, called \emph{pair geometry} (Figure~\ref{fig:dockingSites}b). As the parts are in dense correspondence, we have dense correspondences between all pair geometry that connects the same part type through the same pair of boundaries. We build the probabilistic shape space for the pair geometry by a simple PCA analysis. We add the additional energy to the overall energy for all docked part pairs.

To avoid discontinuities, we use smooth weights for all singleton part and pairwise constraints (Figure~\ref{fig:dockingSites}c): The attraction to the shape spaces of the parts fades continuously to zero when approaching the boundaries. Contrarily, the attraction to the pair geometry model grows when moving towards the boundary of the parts.
We weight each vertex by $\exp(-d^2/\sigma_{bdr}^2)$, where $d$ is the distance to the boundary. For seamless results, we set $\sigma_{bdr}$ to to blend within about one third of the part diameter.

\section{Results}
\label{sec:Results}

\begin{figure*}
  \centering
  \renewcommand{\tabcolsep}{0pt}\noindent\begin{tabular}{cc}
    \raisebox{-0.5\height}{\myincludegraphics[width=0.49\linewidth]
      {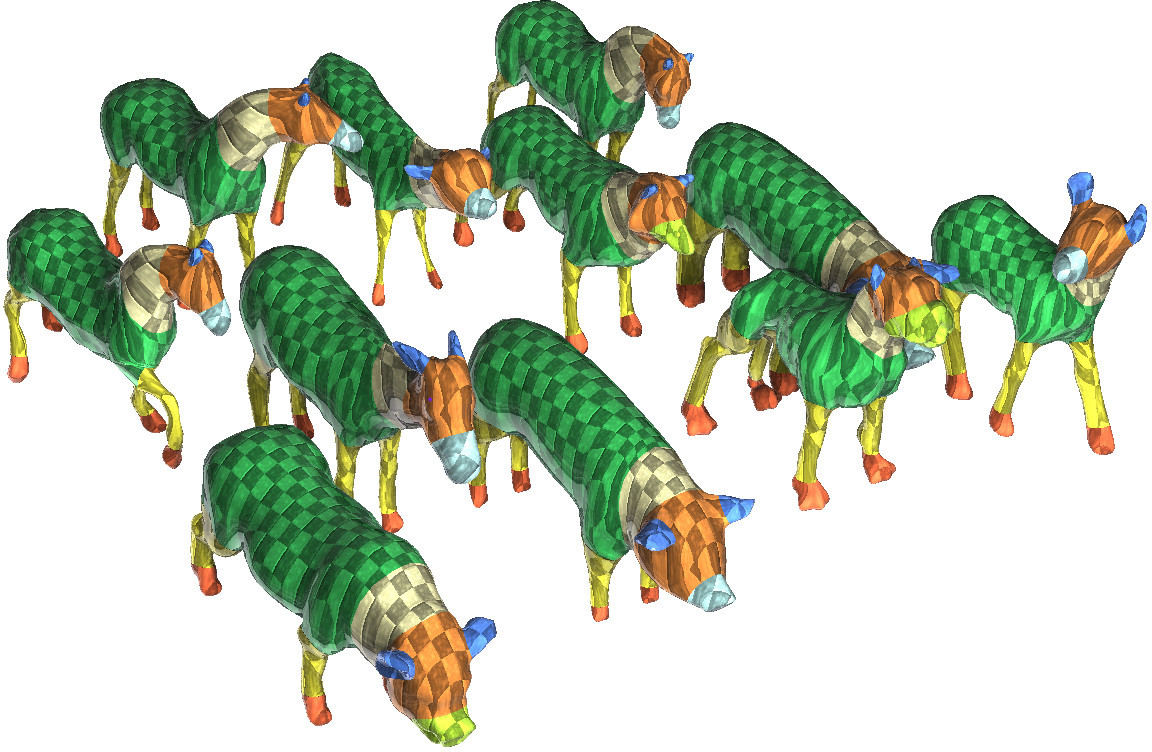}}
    &
    \raisebox{-0.5\height}{\myincludegraphics[width=0.49\linewidth]
      {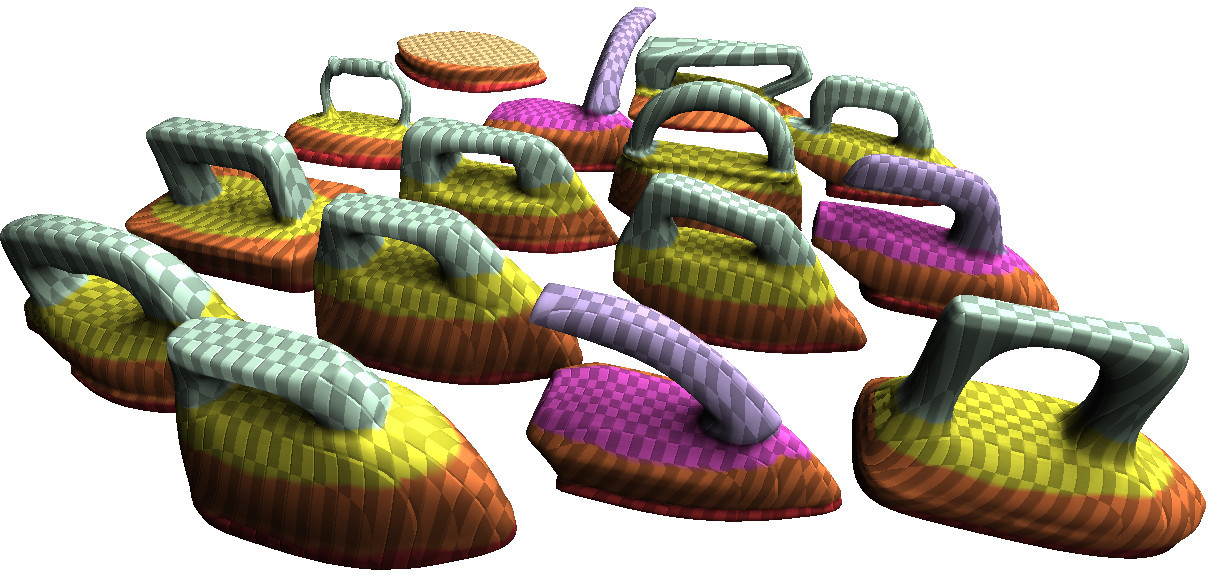}}
    \\
    (a) four legged animals &
    (b) irons
    \\[8pt]
    \raisebox{-0.5\height}{\myincludegraphics[width=0.49\linewidth]
      {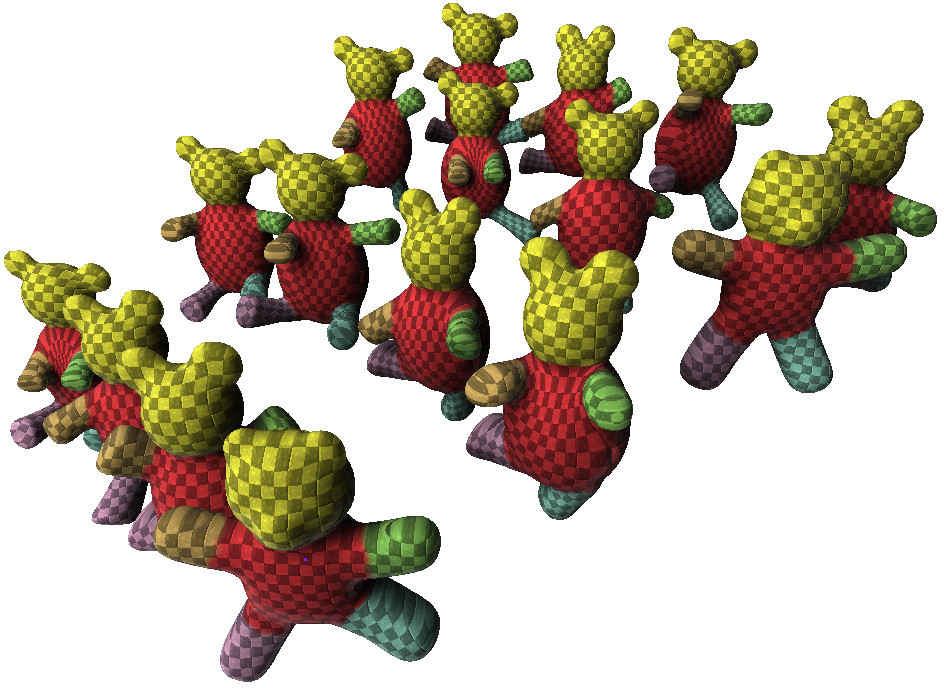}}
    &
    \raisebox{-0.5\height}{\myincludegraphics[width=0.49\linewidth]
      {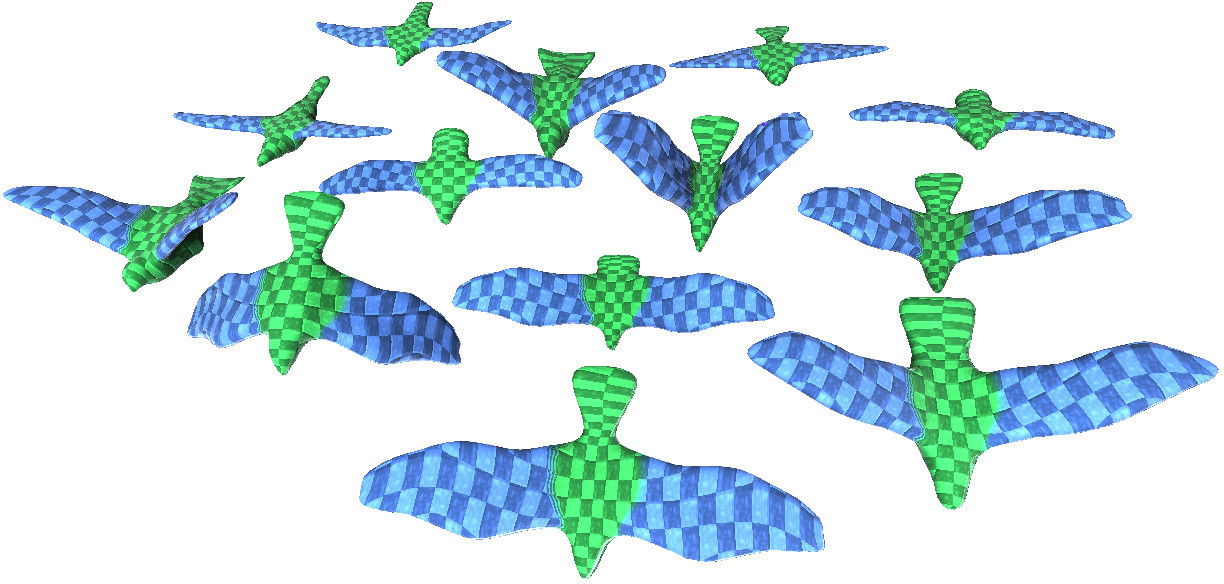}}
    \\
    (c) teddy bears &
    (d) birds
    \\[8pt]
    \raisebox{-0.5\height}{\myincludegraphics[width=0.49\linewidth]
      {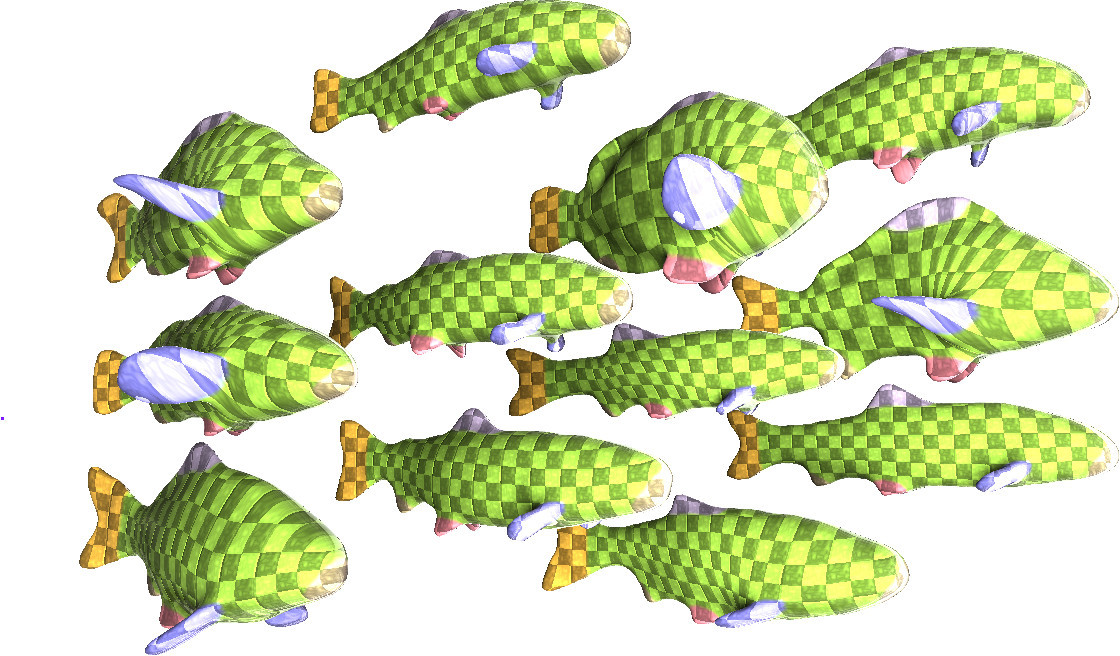}}
    &
    \raisebox{-0.5\height}{\myincludegraphics[width=0.45\linewidth]
      {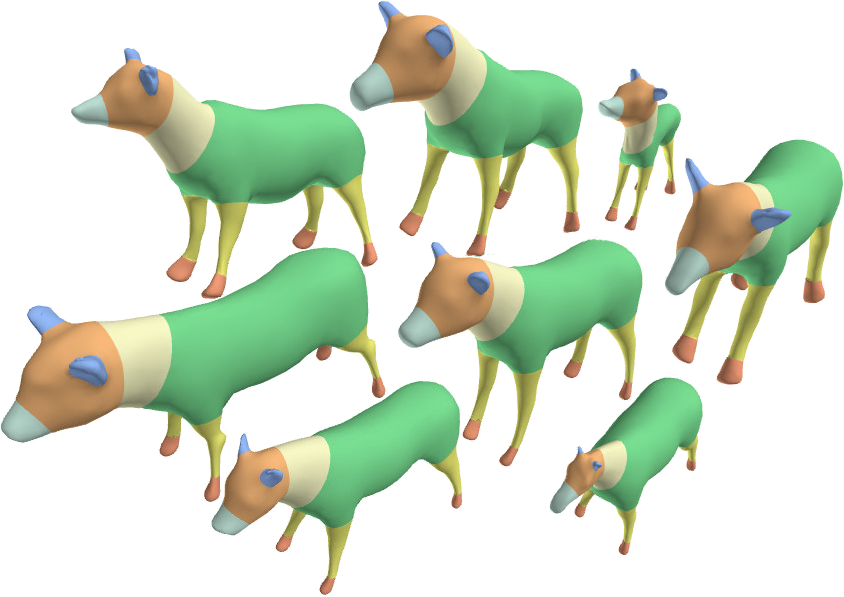}}
    \\
    (e) fish &
    (f) samples from the
    \\
    &
    shape space of (a)
  \end{tabular}
  \caption[Correspondences obtained with our ensemble optimization]{Correspondences obtained with our method for different test data sets taken from the SHREC~2007 benchmark.}
  \label{fig:results}
\end{figure*}

\begin{figure*}
  \centering
  \begin{tabular}{p{.47\linewidth}p{.47\linewidth}}
    \raisebox{-0.5\height}{\myincludegraphics[scale=0.32]
      {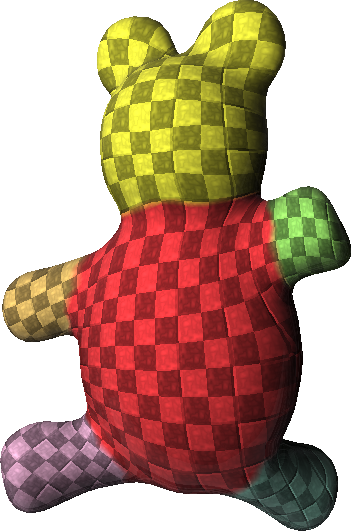}}\ignorespaces
    \raisebox{-0.5\height}{\myincludegraphics[scale=0.32]
      {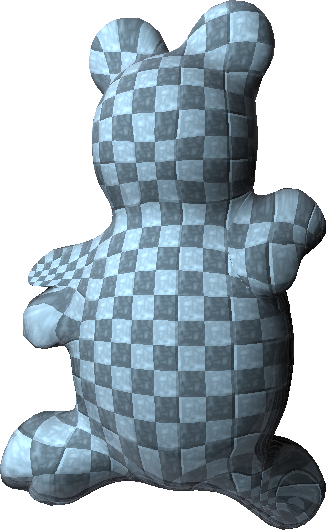}}
    &
    \raisebox{-0.5\height}{\myincludegraphics[scale=0.36]
      {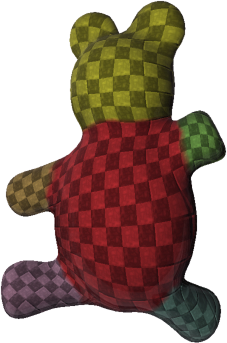}}
    \raisebox{-0.5\height}{\myincludegraphics[scale=0.36]
      {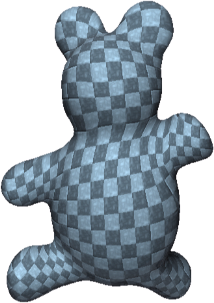}}
    \\
    (a) part-based vs. global optimization high penalty on $E_H$
    &
    (b) part-based vs. global optimization lower penalty on $E_H$
  \end{tabular}
  \begin{tabular}{p{.47\linewidth}}
    \myincludegraphics[width=\linewidth]
    {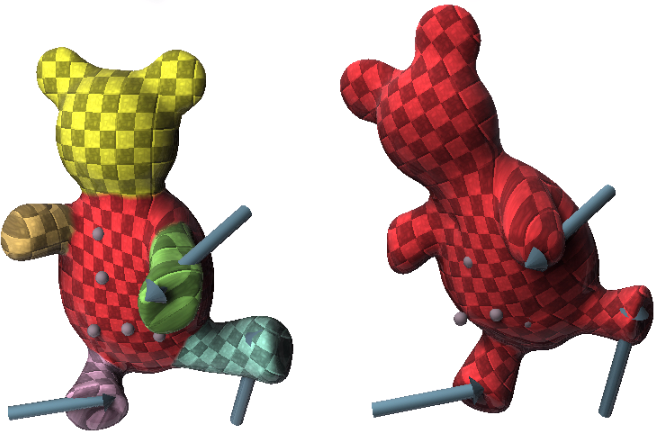}
    \\
    (c) editing:  part-based (left) vs. global (right)
  \end{tabular}
  \caption[Impact of parts]{Impact of parts: (a) For high entropy penalties, a one-part model cannot capture the pose variations, while our method still yields good results. Using a lower weight (higher $\mu_L$; factor 10) resolves the problem, albeit with higher entropy for the global method. (c) shows interactive editing with constraints. Here, the global model has overfit to global pose correlations while parts avoid the effect.}
  \label{fig:results-extra}
\end{figure*}

\begin{figure*}
  \centering
  \renewcommand{\tabcolsep}{0pt}
  \begin{tabular}{ccc}
    \hspace{-.05\textwidth}
    \myincludegraphics[width=0.38\textwidth]
    {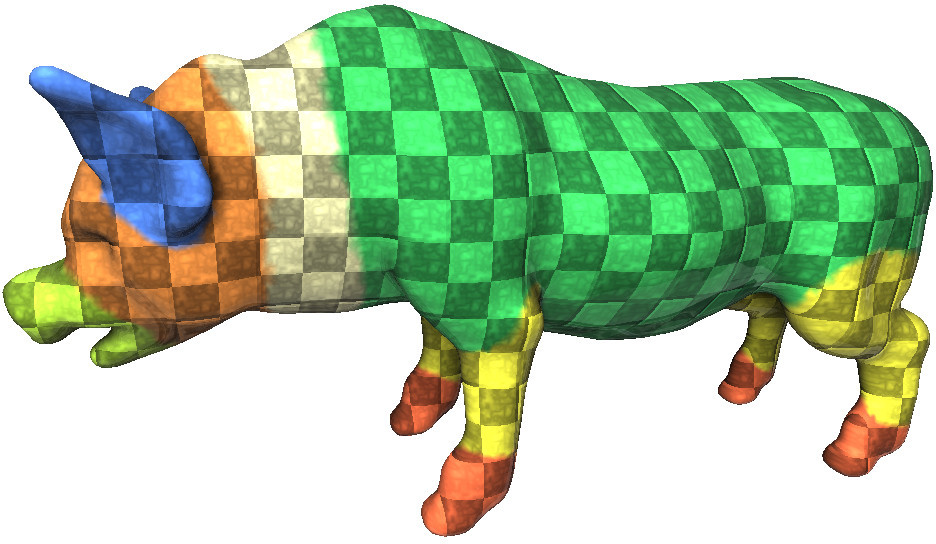} &
    \myincludegraphics[width=0.28\textwidth]
    {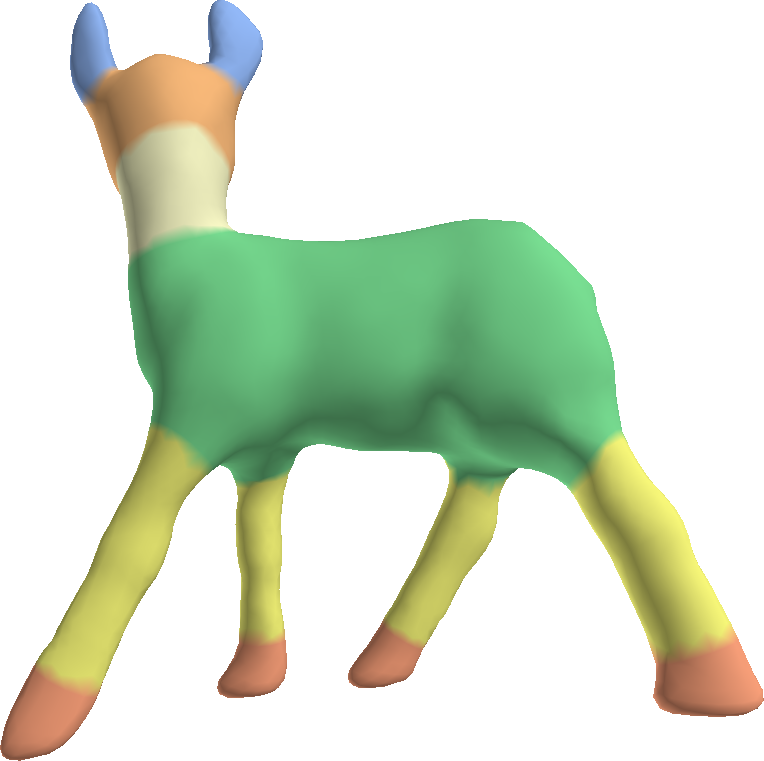} &
    \hspace{-.6em}
    \myincludegraphics[width=0.46\textwidth]
    {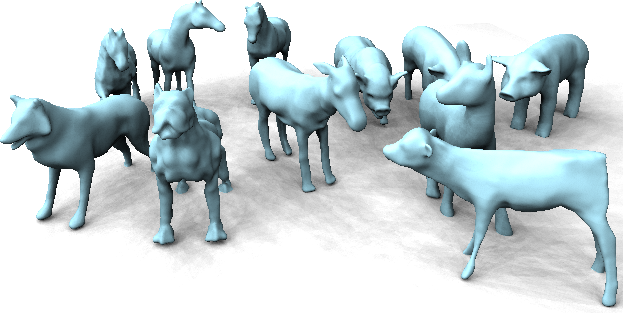}
    \hspace{-.14\textwidth}
    \\
    (a) source shape &
    (b) target shape &
    (c) input shape collection
    \\
    \myincludegraphics[width=0.33\textwidth]
    {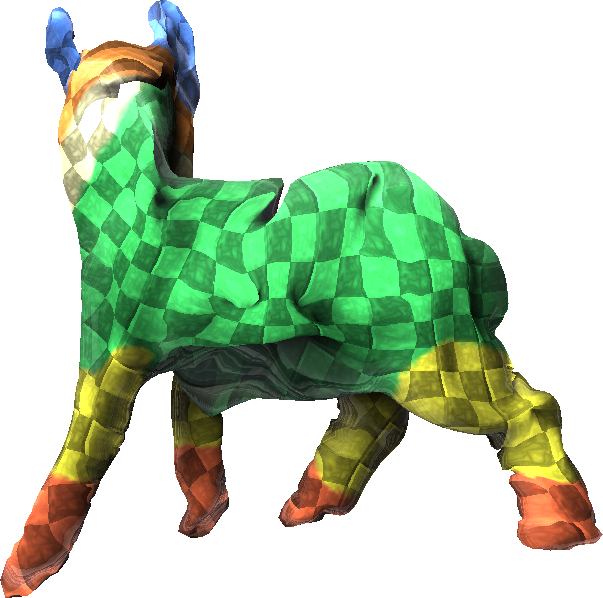} &
    \myincludegraphics[width=0.33\textwidth]
    {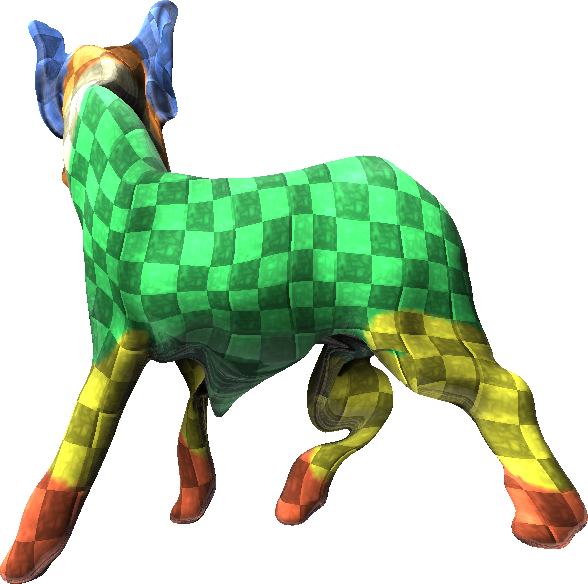} &
    \myincludegraphics[width=0.31\textwidth]
    {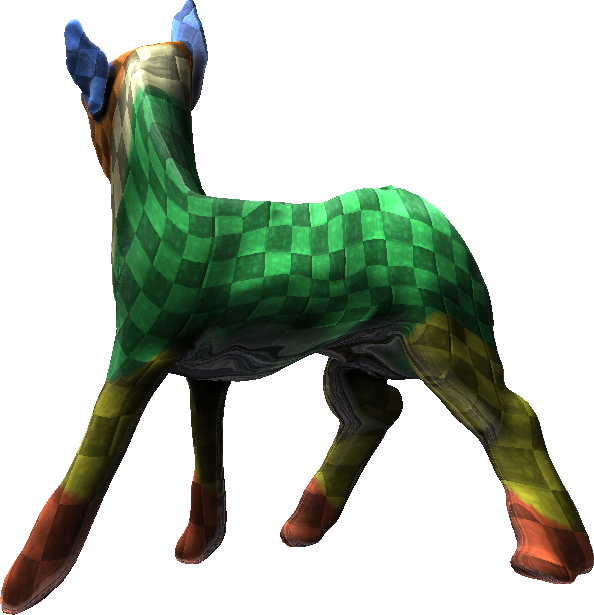}
    \\
    (d) elastic (ARAP) &
    (e) elastic (subspace) &
    (f) thin-plate splines
    \\
    \myincludegraphics[width=0.33\textwidth]
    {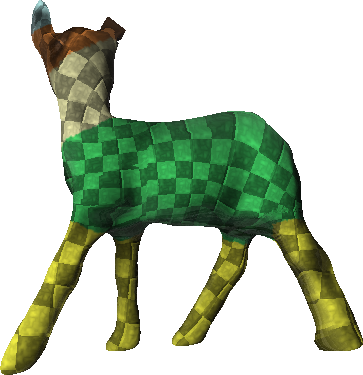} &
    \myincludegraphics[width=0.27\textwidth]
    {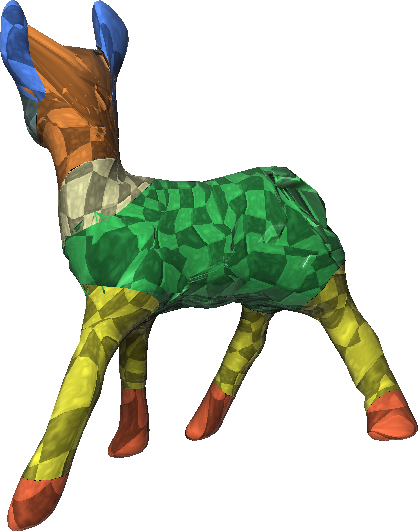} &
    \myincludegraphics[width=0.31\textwidth]
    {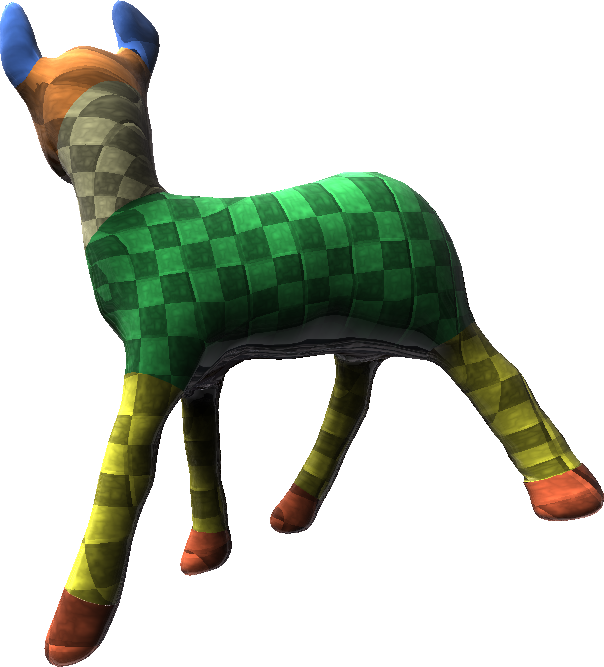}
    \\
    (g) blended intrinsic maps &
    (h) hub and spoke &
    (i) our result
  \end{tabular}
  \caption[Comparison to previous work]{Comparison to previous work. We map source shape (a) to target (b). We use deformable ICP based on as-rigid-as-possible (d) and subspace (e) deformation, as well as (f) thin-plate splines, all initialized with 15 landmarks. We have also employed (g) blended intrinsic maps, (e) ensemble optimization by Huang et al.~[2012]. Our result is shown in (h);     (g) and (h) are ensemble matches, optimized over 19 and 12 animals, respectively, as shown in (i). Please watch the accompanying video to see the impact on the resulting shape spaces; the differences are much more visible in morphing and sampling.}
  \label{fig:results-comparisons}
\end{figure*}

We have implemented the method in C++ and tested the implementation on a dual socket PC (Intel Core i7 with 2.6Ghz and 6 cores per processor). The results are shown in Figures~\ref{fig:teaser} and \ref{fig:results}-\ref{fig:results-comparisons}. We strongly encourage the reader to watch the accompanying \videoWithFootnote, which shows interpolation and sampling results from the constructed shape spaces; these make the improvements due to our method much clearer.

\textbf{Dense correspondences from coarse co-segmentation:} We use the painting interface discussed in Section~\ref{sec:Analysis} to annotate a number of models from the SHREC 2007 model collection. The user has to mark the colored regions shown in Figure~\ref{fig:teaser}b,\ref{fig:results-comparisons}b by  painting on the surface. Additionally, point-to-point correspondences have to be set if the initialization is not clear. For example, for the birds (Figure~\ref{fig:results}d), the tip of the wings needed one more such point match per wing. Additional constraints are not always necessary; for example, the animals data set has been build from the user segmentation only. After such initialization, we run the optimization. The user has to chose the parameter $\mu_L$ as well as the level of resolution for the remeshing step (after initial cross parametrization). The first parameter is critical for the results, the second trades-off run-time and accuracy. Finding an appropriate annotation and parametrization that works for a whole shape ensemble requires multiple iterations of interaction and optimization. Here, computing correspondences, minimizing Eq.~\ref{eq:energy}), took on average about 20min for a shape set. Given the additional steps (parametrization, remeshing etc.), the net computation time adds up to to roughly 1h per model. In our examples, interaction and computation amounted to up to 6h for our example models, depending on the complexity (e.g., animals were more difficult than teddies). It would probably be possible to automate the procedure by using recent fully automatic co-analysis \cite{Huang2011,Sidi2011,Huang2012,Kim2012} for initialization, but this is still subject to future work.

\textbf{Results:} The resulting correspondences within several subsets of this collection are shown in Figure~\ref{fig:results}: We use a checker-board texture projected to one instance of the input and transfer it to all other models to visualize correspondences. Additionally, differently colored regions depict parts. The resulting correspondences capture salient features of the models and there is not unwarranted drift. This is very well visible in our video, where we obtain good interpolations for within all of the shape spaces: Intermediate shapes due to morphing as well as due to random sampling from the underlying Gaussian at the learned standard deviations (sampling at $\sigma=1$) are plausible (Figure~\ref{fig:results}f). We should highlight that the model is able to handle structures with fine details, such as the legs of the animals in Figure~\ref{fig:results}a. As discussed in Section~\ref{sec:Analysis}, we have tried various alternative approaches, all of which failed at this data set.

\textbf{Comparison to pairwise local registration:} We compare to a number of base-line methods first. In all cases, matching is done by first sampling 43 points uniformly from our solution to be used as initialization and then switching to deformable ICP \cite{Allen2003}. We have also tried alignment without landmarks, as well as deactivating landmarks; the show result (keeping the landmarks during ICP) yielded the best results.

Figure~\ref{fig:results-comparisons} shows the results of pairwise matching between a pig and a young deer. We examine as-rigid-as-possible (ARAP) deformation model (\cite{Sorkine2007}, Figure~\ref{fig:results-comparisons}d) and a closely related variant, using a smooth subspace deformation model (\cite{Adams2008}, Figure~\ref{fig:results-comparisons}e): The subspace model uses a volumetric low-frequency basis ($12^3$ grid in our case), which leads to smoother results than ARAP. Nonetheless, both cases suffer from artifacts such as wrinkles and drifting correspondences. The video illustrates the disastrous effect on the obtained shape spaces.

Thin-plate splines (TPS) are substantially better (Figure~\ref{fig:results-comparisons}f), which was to be expected as this is the current standard solution for this type of matching problems \cite{Allen2003,Hasler2009}. Nonetheless, the TPS model still creates wrinkles and unwarranted drift. Our video shows various artifacts in the resulting shape spaces.

\textbf{Intrinsic matching methods:} In order to compare to recent state-of-the-art methods, we have also performed pairwise matches with blended intrinsic maps (BIM) \cite{Kim2011}. As shown in Figure~\ref{fig:results-comparisons}g, the pairwise partial isometries cannot capture the variation in this challenging data set well (please note though, that BIM is a global correspondence method; it solves a more difficult problem than our paper). Subsequently, we use the hub-and-spoke ensemble matcher of Huang et al.~\shortcite{Huang2012} (also a global matcher) that takes these results as input and performs a selection of best partial isometries in order to create consistent equivalence relations. The method improves the quality of the correspondences (Figure~\ref{fig:results-comparisons}h), but substantial misalignment persists, which is seen best in the morphs shown in our video (remark: the output of their method is not as dense as the original vertices; we use interpolation to visualize the results; the same artifacts are also visible in the sparser output alone). In comparison, our method (Figure~\ref{fig:results-comparisons}h) has a very good feature alignment and virtually no drift (see the video).

\textbf{Modeling with deformable parts:} The reconstruction from part graphs is formulated as an optimization problem. A variational approach permits us to easily include additional constraints. For example, any of the points can be fixed. We can for example use an energy $(\vecv_i-\vecy)^2$ to implement handles that the user can attach to the shape for editing. Further, shape parameters can be prescribed. We use energies of the form $(\shapeParam_i - y)^2$ to control the shape of individual parts. We can also couple parameters of different parts (energies of the form $(\shapeParam_i - \shapeParam_j)^2$), for example, to keep shapes of the same type symmetric. The accompanying video shows some morphs between shapes with random shape space parameters, as well as an interactive editing session. A result of interactive editing is also shown in Figure~\ref{fig:results-extra}c.

\textbf{Impact of the part-based model:} Using a part-based approach has a number of advantages: First, as shown in Figure~\ref{fig:results}a,b and in the video, we can capture discrete, topological variations in addition to continuous shape parameters: The irons consist of parts that can be assembled in different variations; the four-legged animals also distinguish open and closed mouths.
Learning these shape families would be very challenging with global approaches. Despite the part-wise approach, our gradient-domain synthesis algorithm yields perfectly smooth boundaries in all cases (deactivating for example the smoothed connections illustrated in Figure~\ref{fig:dockingSites}c degrades the quality significantly).
Further, our learning method benefits from symmetry within a shape; for example, all four legs in each animal share the same shape space, similarly the wings of the birds.
We also obtain additional benefits: As illustrated in Figure~\ref{fig:results-extra}a,b we can learn more compact shape spaces using well-chosen parts. A global models yield bad correspondences for strong entropy penalties (i.e., low values of $\mu_L$). Reducing these improves the results, but the model then learns global correlations that are often unwanted. For example, in the case of the teddy bear in Figure~\ref{fig:results-extra}c, global pose correlations are captured, which tilt the object against the position constraints on the chest; this that make harder and yields worse results (see also the video for an animated visualization). In summary, parts give us a more flexible model that allows us to integrate topologically diverse shapes and to learn shape spaces from fewer examples, avoiding overfitting.

\textbf{Limitations:} The most important limitation of our method is that it is a local optimization technique, thus requiring quite some user interaction as well as parameter choices. Although we require only a coarse initialization, a too coarse annotation causes the algorithm to get stuck in a local optimum. A further, theoretical limitation is that we cannot formally guarantee bijectivity of correspondences, but we have not observed problems in practice. Finally, the hard-constraints for the surface constraints in the optimization limits the applicability to manifold input. Noisy and, in particular, incomplete data from 3D scans currently cannot be handled.

\section{Conclusions and Future Work}
\label{sec:ConclusionsAndFutureWork}

We have presented a new method for refining correspondences in families of shapes. By taking the compactness of the shape space into account as an optimization criterion, we obtain high-quality dense correspondences well-suited for the creation of shape spaces among shapes of considerable variability.  In direct comparison, previous methods show substantial artifacts in such situations that we can avoid. Even difficult situations such as strong deformations and widely varying geometry yield good results. Our method handles objects of general topology, it handles challenging meshes with small feature sizes reliably, and is able to learn from objects of varying part composition, which can be used to synthesize new shapes with variable part configuration and continuous variability that adapts automatically to the designed part layout. Further, the part-based approach yields higher quality correspondences and is a useful tool to avoid overfitting.

In future work, we would like to extend the method towards fully automatic global matching, avoiding tedious manual initialization. Recent progress in co-segmentation would provide a starting point here, but a fully automatic method would require making our method robust to slight variations in part topology and outlier mismatches. In the long term, the question of how to build compact explanations from observed data is of fundamental importance. An ultimate modeling system with deformable parts would decompose shape collections automatically to obtain a shape grammar and various deformable, dockable shape spaces of parts, both optimized for compactness of encoding. While our model can in principle already handle such scenarios in terms of representation and synthesis, the automated analysis is the key challenge.

\bibliographystyle{alpha}
\bibliography{entcor}

\end{document}